\documentclass[11pt]{article}
\usepackage{epsf,amsmath,amssymb}
\usepackage[nosort]{cite}
\newdimen\SaveWidth \SaveWidth=\textwidth
\newdimen\SaveHeight \SaveHeight=\textheight
\advance\SaveWidth by -\textwidth
\advance\SaveHeight by -\textheight
\divide\SaveWidth by 2
\divide\SaveHeight by 2
\advance\hoffset by \SaveWidth
\advance\voffset by \SaveHeight
\newcommand{\code}{\small\sf}
\newcommand{\abbrev}{\small}
\newcommand{\ep}{\epsilon}
\newcommand{\eqn}[1]{Eq.\,(\ref{#1})}
\newcommand{\Eqn}[1]{Equation~(\ref{#1})}
\newcommand{\fig}[1]{Fig.\,\ref{#1}}
\newcommand{\Fig}[1]{Figure~\ref{#1}}

\newcommand{\sct}[1]{Sect.\,\ref{#1}}
\newcommand{\dd}{{\rm d}}
\newcommand{\ddoverdd}[1]{{\dd\over \dd #1}}

\newcommand{\order}[1]{{\cal O}(#1)}

\newcommand{\bare}{{\rm B}}
\newcommand{\api}{{\alpha_s\over \pi}}

\newcommand{\apib}{{\alpha_s^\bare\over \pi}}

\newcommand{\msbar}{\overline{\mbox{\small MS}}}
\newcommand{\Lx}{\left(}
\newcommand{\Rx}{\right)}
\newcommand{\LB}{\left[}
\newcommand{\RB}{\right]}
\newcommand{\issoft}{\,{\buildrel{\rm soft}\over\longrightarrow}\,}
\newcommand{\intdy}{\int_0^1\kern-5pt dy}
\newcommand{\intdz}{\int_0^1\kern-5pt dz}

\newcommand{\logmuh}[1]{l_{H}^{#1}}
\newcommand{\logmut}[1]{l_{t}^{#1}}
\newcommand{\pdf}{{\abbrev PDF}}
\newcommand{\qcd}{{\abbrev QCD}}
\newcommand{\SM}{{\abbrev SM}}
\newcommand{\lo}{{\abbrev LO}}
\newcommand{\nlo}{{\abbrev NLO}}
\newcommand{\nnlo}{{\abbrev NNLO}}
\newcommand{\lep}{{\abbrev LEP}}
\newcommand{\lhc}{{\abbrev LHC}}

\title{
  \vspace{-3em}
  \begin{flushright}
    \sf\normalsize BNL-HET-01/6,
    \sf\normalsize hep-ph/0102241 ---
    \sf\normalsize 19 February 2001\\[3em]
  \end{flushright}
  Soft and virtual corrections to $pp\to H + X$ at 
  next-to-next-to-leading order}
\author{Robert V. Harlander\footnote{email address: {\tt
  rharlan@bnl.gov}}\ \ and
  William B. Kilgore\footnote{email address: {\tt kilgore@bnl.gov}}\\[1em]
  \it\normalsize HET, Physics Department\\
  \it\normalsize Brookhaven National Laboratory, Upton, NY 11973, U.S.A.
  }
\date{}
\begin{document}
\maketitle {\abstract The contributions of virtual corrections and soft
  gluon emission to the inclusive Higgs boson production cross section $pp\to
  H + X$ are computed at next-to-next-to-leading order in the heavy top
  quark limit. We show that this part of the total cross section is
  well behaved in the sense of perturbative convergence, with the
  {\abbrev NNLO} corrections amounting to an enhancement of the
  {\abbrev NLO} cross section by $\sim5$\% for the CERN {\abbrev LHC} and
  10-20\% for the Fermilab Tevatron.  
  We compare our results with an existing estimate of the full
  {\abbrev NNLO} effects and argue that an analytic evaluation of the
  hard scattering contributions is needed.
  }



\section{Introduction}

In the past two decades the predictions of the standard model have been
confirmed with remarkable precision.  Still, the agent of electroweak
symmetry breaking remains elusive.  The simplest mechanism introduces a
single complex scalar doublet.  Three components of the doublet give
mass to the $W$ and $Z$ bosons, leaving a single neutral particle, the
Higgs boson, as the signature of the symmetry breaking sector.  This
scenario is called the minimal standard model (\SM) and forms the
benchmark for the investigation of electroweak symmetry breaking.

In their final run, the experiments at the CERN $e^+e^-$ collider
\lep\ established a lower mass limit for a \SM\ Higgs boson of $\sim113$
GeV~\cite{LEPHiggs}.  This value severely challenges the reach of the
Fermilab Tevatron.  Nonetheless, there are indications that the Higgs
boson is not
much heavier than this limit.  Fits to precision electroweak data
actually prefer a value of the Higgs boson mass that is well below the
exclusion limit and place a $95\%$ confidence level upper limit near
$200$\,GeV~\cite{LEPSMFit}.  Supersymmetric extensions to the \SM,
which have at least two Higgs doublets and therefore at least five
observable Higgs bosons, prefer that the mass of the lightest Higgs be
below approximately $135$~GeV~\cite{SUSYHiggs}.  Detailed studies
indicate that, with sufficient luminosity, the Tevatron could be
sensitive to Higgs bosons somewhat beyond the $W^+W^-$
threshold~\cite{HTZ}.

At hadron colliders, the dominant production mechanism for Higgs bosons
of mass below $\sim700$\,GeV is gluon fusion.  The Tevatron however will
only be able to make limited use of this production channel.  For
Higgs boson
masses well above the $W^+W^-$ threshold, the production rate is too
small to be observed.  For masses well below the $W^+W^-$ threshold, the
dominant $b\bar{b}$ decay mode is overwhelmed by \qcd\ background, and
the production rate is insufficient to allow observation of rare decay
modes like $H\to\gamma\gamma$.  Thus except for a 
window around
the $W^+W^-$ threshold where that decay mode may be observable, the
Tevatron Higgs search will rely on associated production with a $W$ or
$Z$ boson rather than gluon fusion~\cite{TevRept}.

At the CERN Large Hadron Collider (\lhc) the gluon fusion production mechanism will be extremely
important.  For lighter Higgs boson masses, the $b\bar{b}$ decay mode will
still be overwhelmed by \qcd\ effects, but because of the high machine
luminosity and the large gluon luminosity at relatively small parton
energy fractions, the \lhc\ will be able to measure the
$H\to\gamma\gamma$ decay mode.  This will be important not only as a
discovery channel (if the Higgs boson has not yet been found) but also as a
means of studying Higgs properties like its coupling to top quarks.

Unfortunately, the gluon fusion process is currently not fully under control.
The next-to-leading order (\nlo) corrections to
the production rate were computed ten years ago and were found to be
extremely large, of order $70-100\%$~\cite{dawdsz,oneexact}.  Such
large corrections clearly ask for the evaluation of higher order terms
in the perturbative series in order to arrive at a solid theoretical
understanding of the process.  In this paper, we present the soft
plus virtual corrections to inclusive Higgs production at
next-to-next-to-leading order (\nnlo).  For the relevant values
of $M_H \sim 100$--$200$\,GeV and the center-of-mass (c.m.s.)
energies at Tevatron (2\,GeV) and LHC (14\,TeV), these terms are not
expected to dominate 
the full result.  Nevertheless, they represent a first step towards
the complete answer, in the sense that they comprise a well-defined,
ultraviolet and infrared finite, gauge and renormalization group
invariant piece. By comparing to existing estimates on the \nnlo\ 
corrections for $gg\to H$, we conclude that the evaluation of the full
answer is necessary.


\section{Higgs production through light parton scattering}


\subsection{Leading order}
\begin{figure}
  \begin{center}
    \leavevmode
    \begin{tabular}{cc}
      \epsfxsize=4.cm
      \epsffile[120 240 510 480]{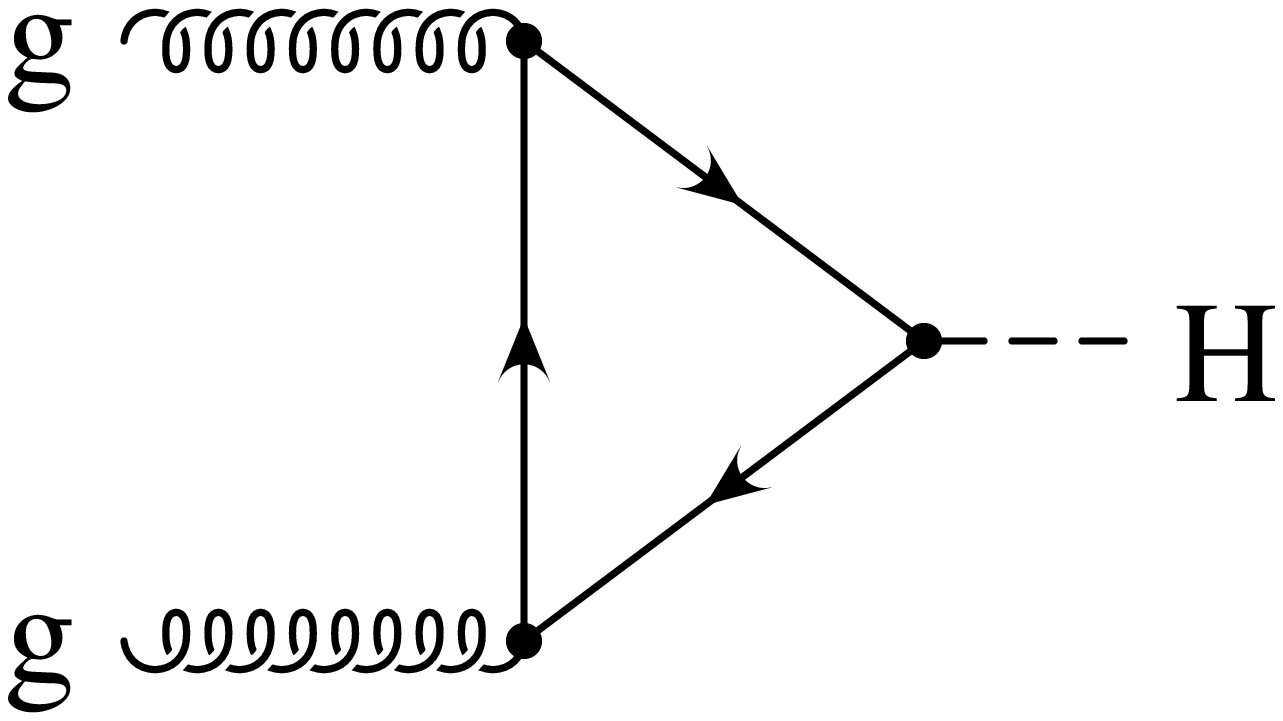} &
      \epsfxsize=4.cm 
      \epsffile[120 240 510 480]{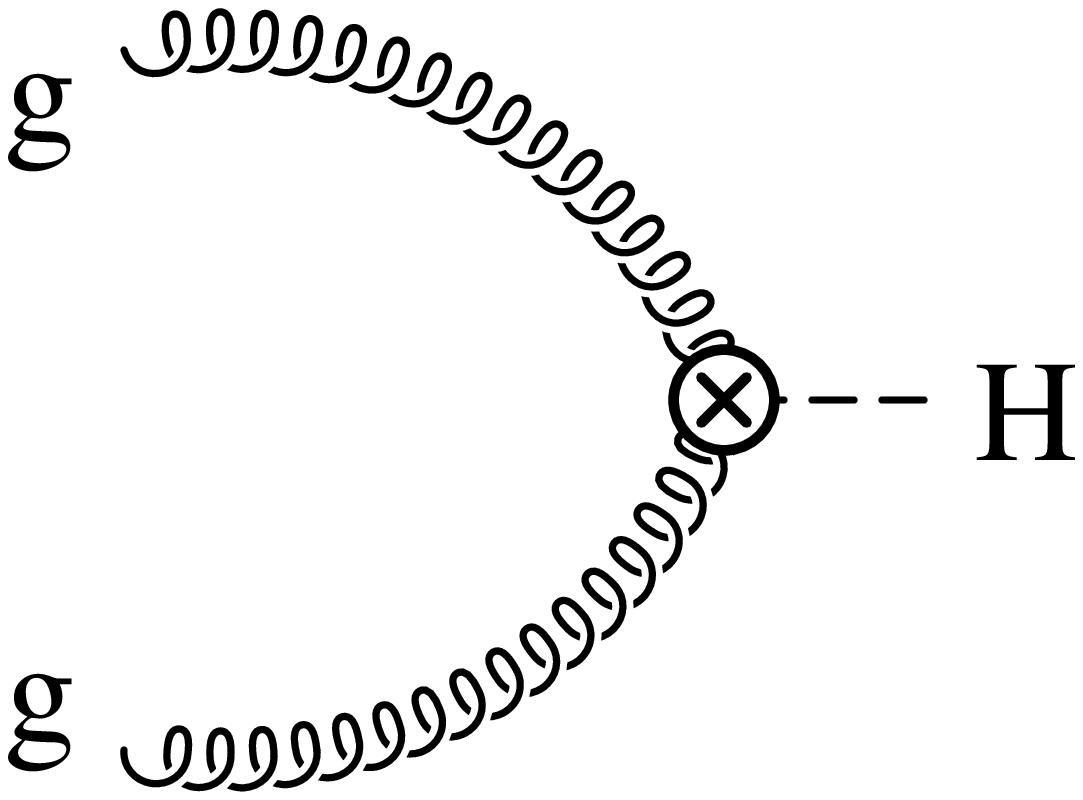}\\
      (a) & (b)
    \end{tabular}
    \parbox{14.cm}{
      \caption[]{\label{fig::lo}\sloppy
        Leading order diagram to the process $gg\to H$: (a)~in full
        \qcd; (b)~in the effective theory [\eqn{eqn::leff}].
        The straight solid lines represent the top quark, the symbol
        $\otimes$ denotes the effective vertex.}}
  \end{center}
\end{figure}

Assuming the standard model as the theory of particle interactions, the
coupling of the Higgs boson to gluons is mediated by a quark loop.
In the following we will neglect all quark masses except for the top
mass. In this case only the top quark contributes, because the Yukawa
couplings are proportional to the fermion masses. The lowest order
diagram is then shown in Fig.\,\ref{fig::lo}\,(a).  It was computed some
time ago~\cite{lo} and leads to the cross section
\begin{equation}
\begin{split}
\sigma_{\rm LO}(gg\to H)
&= {G_{\rm F}\alpha_s^2(\mu^2)\over 128\sqrt{2}\pi}\tau^2\delta(1-x)
\left| 1+(1-\tau)f(\tau) \right|^2\,,\\
f(\tau) &= \left\{
  \begin{array}{ll}
    \arcsin^2{1\over\sqrt{\tau}}\,,& \tau\geq 1,\\
    -{1\over 4}\left[\ln{1+\sqrt{1-\tau}\over 1-\sqrt{1-\tau}} -
      i\pi\right]^2\,, & \tau< 1,
  \end{array}
  \right.\\
  \tau &= 4M_t^2/M_H^2\,, \qquad x = M_H^2/s\,,
\label{eqn::lo}
\end{split}
\end{equation}
where $s$ is the partonic center-of-mass energy, $G_{\rm F}$ the
Fermi coupling constant, $\alpha_s$ the strong coupling constant
(which depends on the renormalization scale $\mu$), $M_t$ the pole
mass of the top quark, and $M_H$ the Higgs boson mass.

\subsection{Effective Lagrangian}
The lowest threshold of the diagram in \fig{fig::lo}\,(a) is at
$M_H = 2M_t$.  This means that an expansion in terms of $M_H/M_t$ is
expected to converge for $M_H< 2M_t$.  In fact, it has been shown that
the \nlo\ K-factor for this process is excellently
approximated up to much larger Higgs boson masses, even by keeping only the
first term of such an expansion.  With a top mass of around $M_t =
175$\,GeV and the experimental data favoring a relatively low Higgs boson
mass between 100 to 200\,GeV, we feel that keeping only the leading
term in $M_H/M_t$ is also well justified at \nnlo.

We therefore integrate out the top quark and compute amplitudes using
\qcd\ with five active flavors and the following effective
Lagrangian~\cite{efflag} for the Higgs-gluon interaction:
\begin{equation}
\begin{split}
{\cal L}_{\rm eff} 
&= -{H\over 4v} C_1\,{\cal O}_1
= -{H\over 4v} C_1^\bare\,{\cal O}_1^\bare,
\qquad\qquad
    {\cal O}_1 =  G^a_{\mu\nu} G^{a\,\mu\nu}\,,
\label{eqn::leff}
\end{split}
\end{equation}
where $G^a_{\mu\nu}$ is the gluon field strength tensor.  In the
approximation that all light flavors are massless, this effective
Lagrangian is renormalization group invariant, but the coefficient
function $C_1^\bare$ and the operator ${\cal O}_1^\bare$ must each be
renormalized.  Below, we give the renormalized value of $C_1$ while
the renormalization of ${\cal O}_1^\bare$ and its matrix elements is
discussed in \sct{sec::renorm}.

The coefficient function contains the residual logarithmic dependence
on the top quark mass and has been computed up to
$\order{\alpha_s^4}$~\cite{CheKniSte97}.  For our purposes, we need it
only up to $\order{\alpha_s^3}$~\cite{CheKniSte97,gghresum}.  The
coefficient function in the modified minimal subtraction scheme
($\msbar$) is
\begin{equation}
\begin{split}
C_1 &= -{1\over 3}\api\bigg\{
1 + {11\over 4}\api 
+ \left(\api\right)^2\left[{2777\over 288} + {19\over 16}\logmut{}
+ n_f\left(-{67\over 96} + {1\over 3}\logmut{}\right)\right]
+ \ldots\bigg\}\,,
\label{eq::c1}
\end{split}
\end{equation}
where $\logmut{} = \ln(\mu^2/M_t^2)$ and $M_t$ is the on-shell top quark
mass.  $\alpha_s \equiv \alpha_s^{(5)}(\mu^2)$ is the $\msbar$
renormalized \qcd\ coupling constant for five active flavors, and
$n_f$ is the number of massless flavors.  In our numerical results, we
always set $n_f = 5$.

Using the effective Lagrangian of \eqn{eqn::leff}, the lowest order
amplitude for the process $gg\to H$ reduces from the one-loop diagram in
\fig{fig::lo}\,(a) to the tree-level diagram in \fig{fig::lo}\,(b).


\subsection{Sub-Processes: virtual, single real, double real}
\label{sec::subprocesses}
For a complete treatment of the process $pp\to H$ in \nnlo\ 
\qcd\ one needs to compute the partonic processes which must
then be convoluted with parton distribution functions.

\begin{figure}
  \begin{center}
    \leavevmode
    \begin{tabular}{ccc}
      \epsfxsize=4.cm
      \epsffile[85 235 490 490]{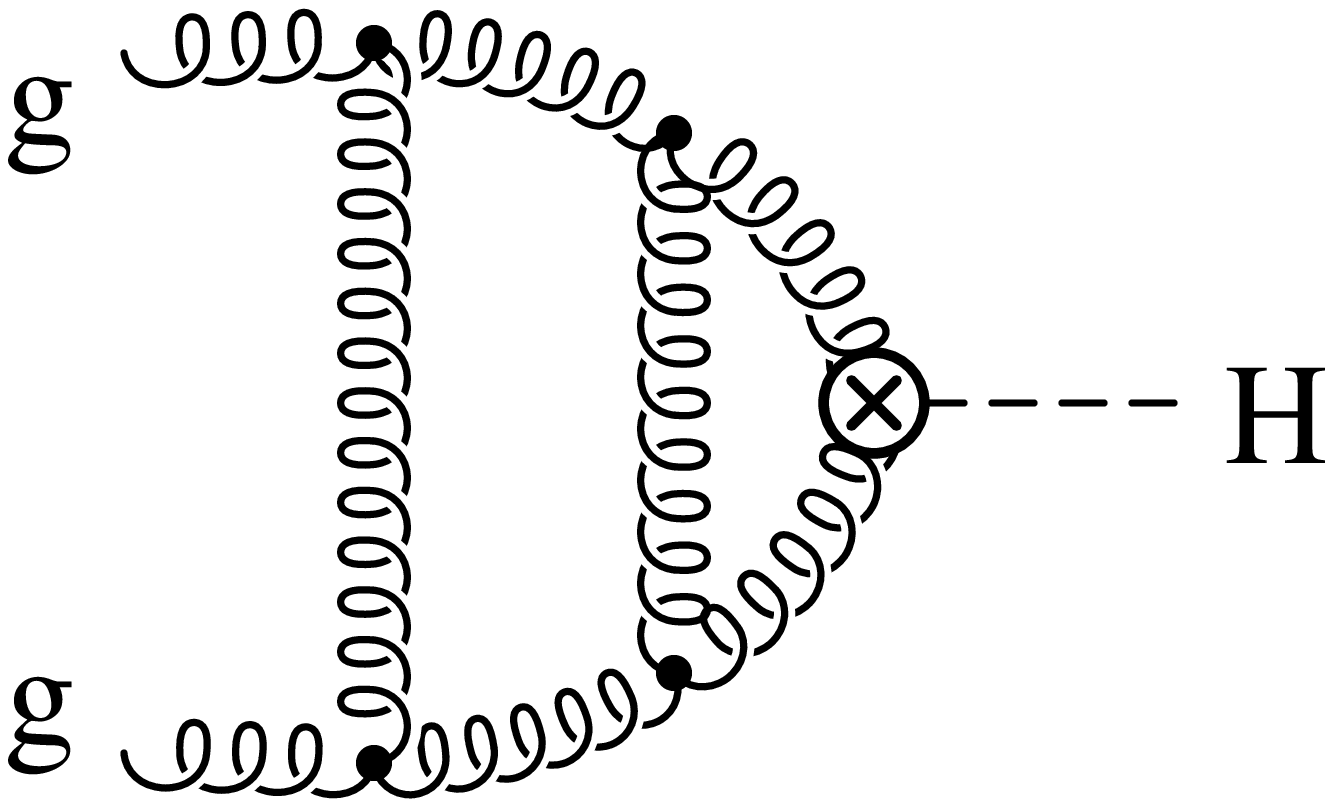} &
      \epsfxsize=4.cm
      \epsffile[85 235 490 490]{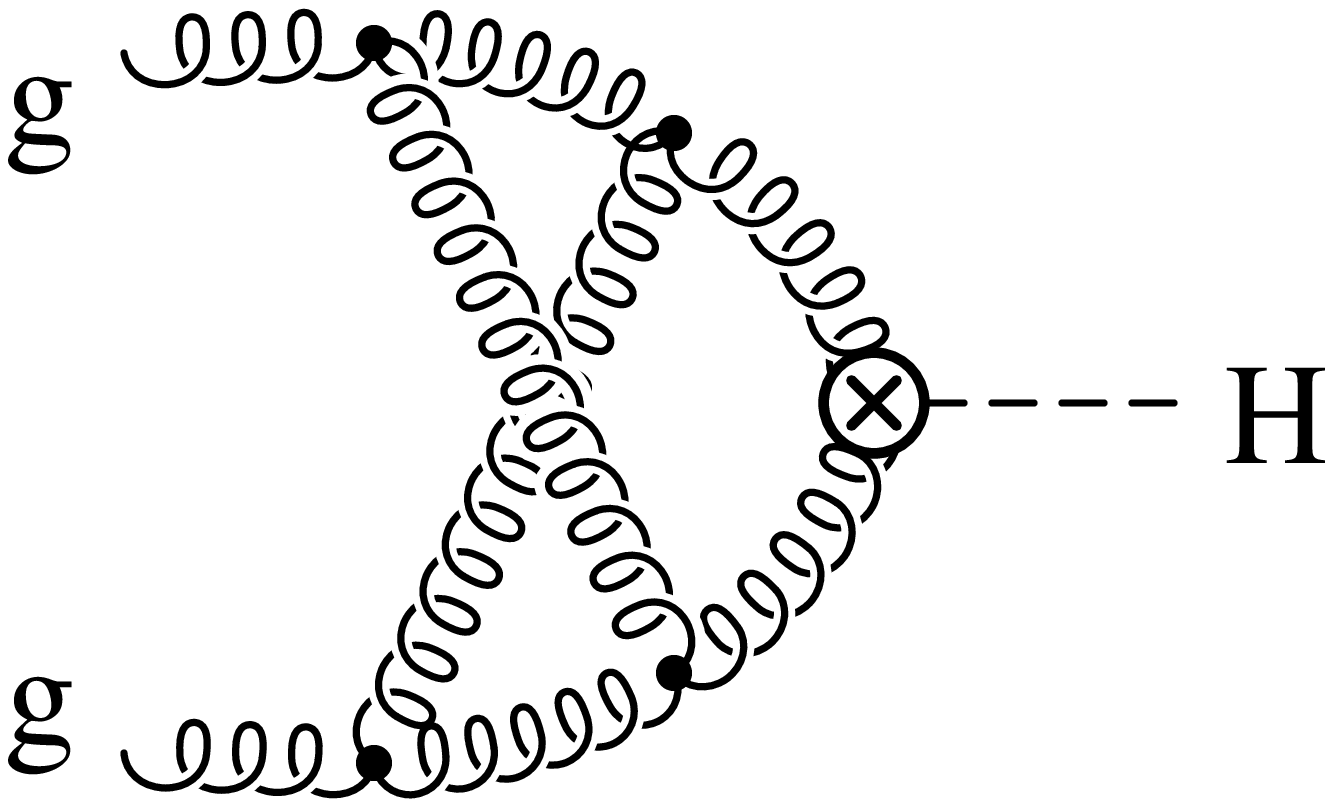} &
      \epsfxsize=4.cm
      \epsffile[85 235 490 490]{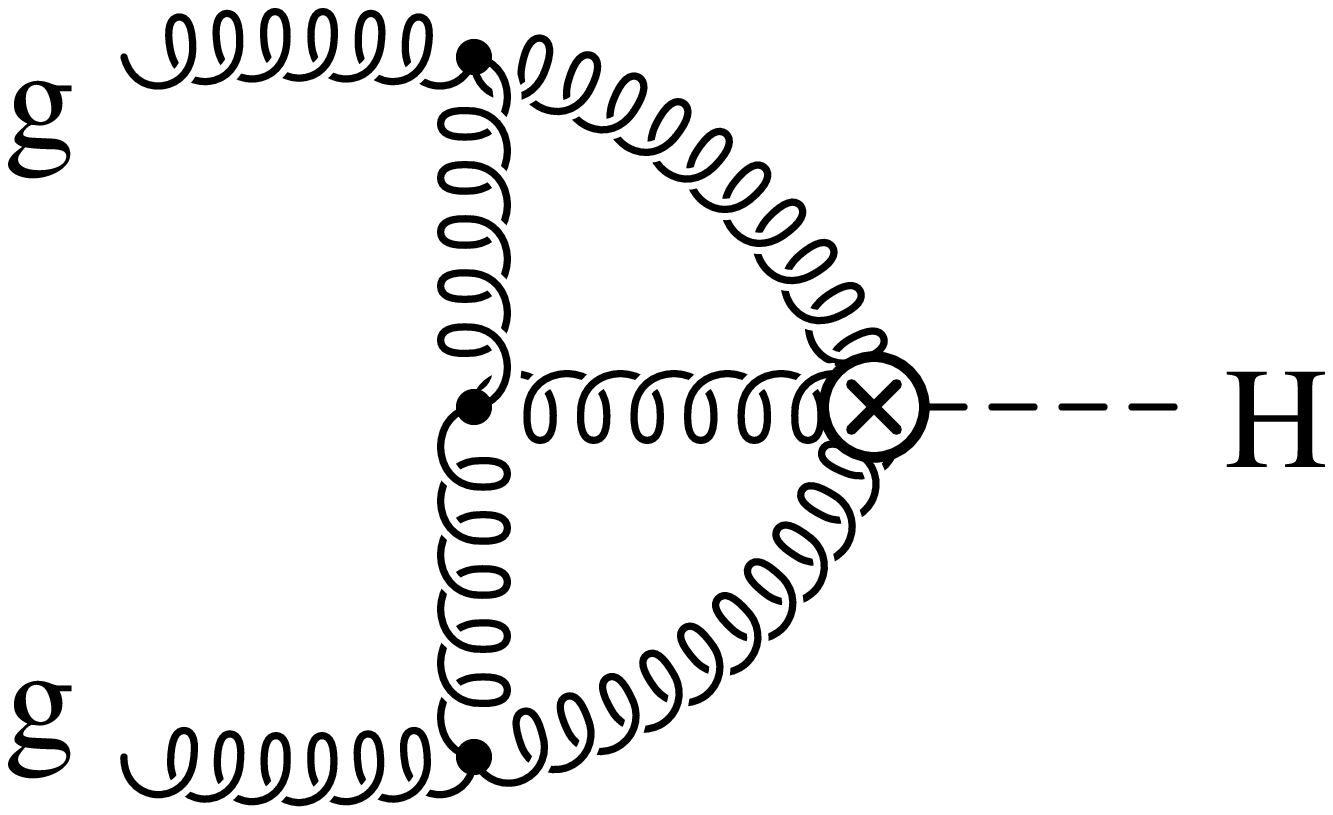} \\
      (a) & (b) & (c)
    \end{tabular}
    \parbox{14.cm}{
      \caption[]{\label{fig::twoloop}\sloppy
        Sample two-loop diagrams contributing to the virtual corrections
        at \nnlo.  }}
  \end{center}
\end{figure}
\begin{figure}
  \begin{center}
    \leavevmode
    \begin{tabular}{ccc}
      \epsfxsize=4.cm
      \epsffile[140 340 440 490]{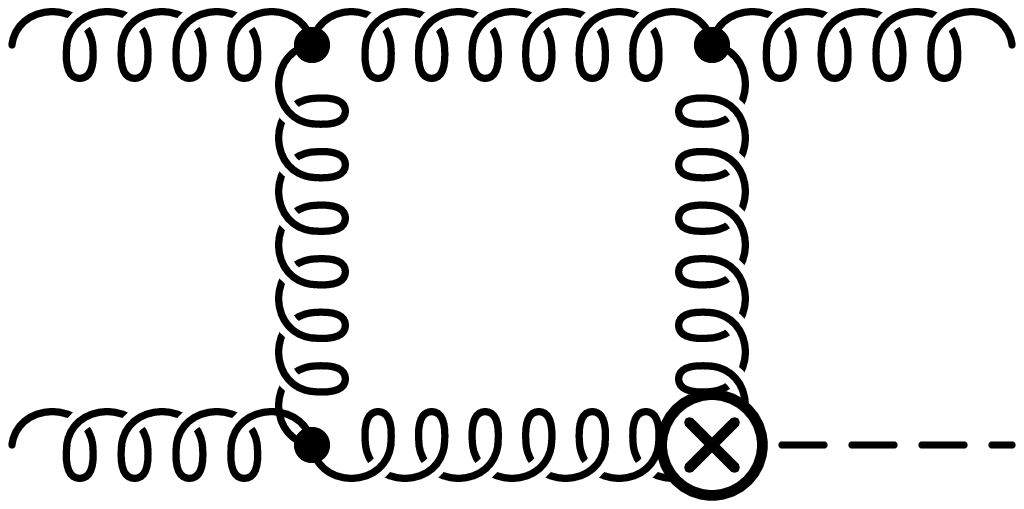} &
      \epsfxsize=4.cm
      \epsffile[140 340 440 490]{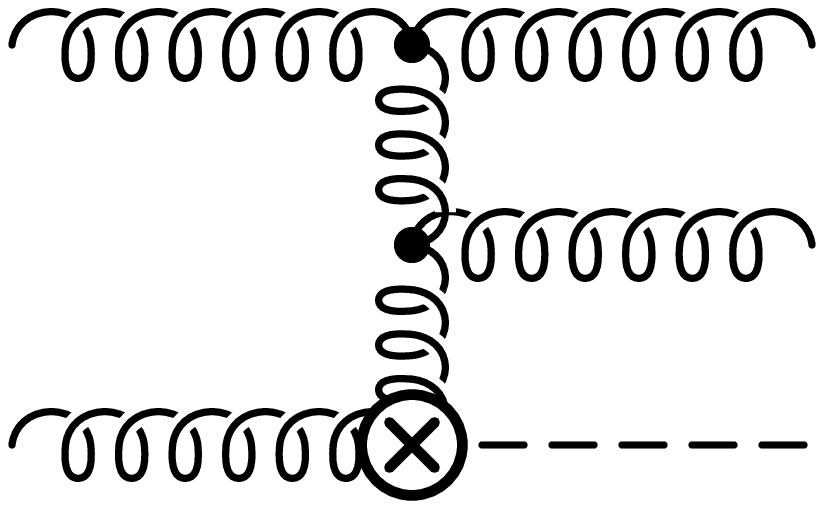} &
      \epsfxsize=4.cm
      \epsffile[140 340 440 490]{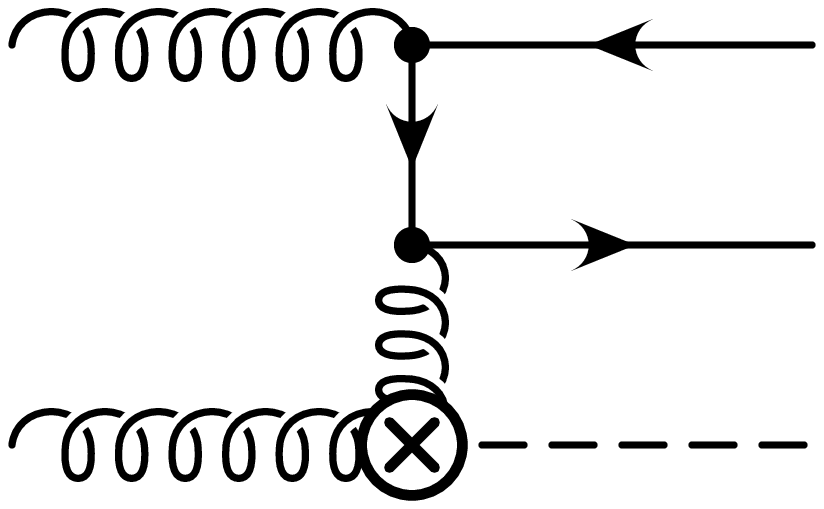} \\
      (a) & (b) & (c)
    \end{tabular}
    \parbox{14.cm}{
      \caption[]{\label{fig::diagrams}\sloppy
        Typical diagrams contributing to (a) single and (b) double real
        radiation of gluons and (c) double real radiation of
        quark-antiquark pairs at \nnlo. The dashed line is the Higgs
        boson and ``$\otimes$'' denotes the effective Higgs-gluon vertex.
        }}
  \end{center}
\end{figure}

The partonic subprocesses are:
\renewcommand{\labelenumi}{($\roman{enumi}$)}
\begin{enumerate}
\item Virtual corrections to two loops:\\
\begin{tabular}{l}
$gg\to H$.
\end{tabular}
\item Single real radiation to one loop:\\
\begin{tabular}{llll}
 $gg\to Hg$\,, &
 $q\bar q\to H g$\,, &
 $qg \to H q$\,, &
 $\bar qg \to H \bar q$.
\end{tabular}
\item Double real radiation at the tree level:\\
\begin{tabular}{llllll}
 $gg\to Hgg$\,,  &
 $gg\to Hq\bar q$\,,  &
 $q\bar q\to Hgg$\,,  &
 $q\bar q\to Hq\bar q$\,,  &
 $qg\to Hqg$\,,  &
 $\bar q g\to H\bar qg$.
\end{tabular}
\end{enumerate}


\subsection{The soft approximation}
\label{sec::softapprox}
Radiative corrections to inclusive production fall into three
categories: delta function terms $\propto\delta(1-x)$, large logarithms
of the form $\ln^n(1-x)/(1-x)$, and hard scattering terms that have at
most a logarithmic singularity as $x\to 1$, where $x$ is the ratio of
the Higgs boson mass squared to the incoming parton c.m.s. energy
squared [see \eqn{eqn::lo}].  In the kinematically similar
Drell-Yan process~\cite{DYsoft} and in inclusive Higgs production at
\nlo~\cite{dawdsz}, it has been observed that the delta function and
large log terms are important though not necessarily dominant parts
of the radiative corrections.

Virtual corrections add only delta function contributions while real
emission corrections contribute to all three categories.  The soft
approximation is obtained by taking the limit $x\to1$.  In doing so, one
is left with only those terms that scale like $(1-x)^{-1-a\ep}$, where
$\ep=(4-d)/2$, and $d$ is the number of space-time dimensions assumed
for the regularization of divergent integrals.  The term
$(1-x)^{-1-a\ep}$ must be interpreted as a distribution,
\begin{equation}
\begin{split}
\label{eqn::distrdef}
f_{a\ep}(x) &\equiv -a\ep\,(1-x)^{-1-a\ep} = \delta(1-x) + \sum_{n=0}^\infty
   {(-a\ep)^{n+1}\over n!}{\cal D}_n(x)\,,
\end{split}
\end{equation}
where ${\cal D}_n(x)$ is a ``plus'' distribution,
\begin{equation}
\begin{split}
{\cal D}_n(x) &= \LB{\ln^n(1-x)\over1-x}\RB_+,\\
\int_0^1\!dx\,h(x){\cal D}_n(x) &= \int_0^1\!dx
  \left[ h(x)-h(1)\right] {\ln^n(1-x)\over1-x}.
\end{split}
\end{equation}
Thus, towers of the large logarithm terms ${\cal D}_n(x)$ can be
trivially obtained from knowledge of the $\delta(1-x)$ terms.



\section{Results}


\subsection{Renormalization of matrix elements}
\label{sec::renorm}
We write the partonic cross section as follows:
\begin{equation}
\begin{split}
  \sigma &= {\pi\over 64 v^2}C_1^2(\alpha_s) Z^2_1(\alpha_s){1\over
  (1-\ep)^2}\left[ \rho_A + \rho_B + \rho_C \right]\,.
\label{eq::sigrho}
\end{split}
\end{equation}
$\rho_A$ is the contribution from virtual corrections, while $\rho_B$
and $\rho_C$ represent the contributions from single and double real
emission, respectively.
$Z_1(\alpha_s)$ is the global renormalization factor for the
composite operator in the effective Lagrangian~\cite{spichet}:
\begin{equation}
\begin{split}
{\cal O}_1 &= Z_1(\alpha_s){\cal O}_1^\bare\,,\\
\mbox{with}\quad Z_1(\alpha_s) &= {1\over 1-\beta(\alpha_s)/\ep}
= 1 - \api\,{\beta_0 \over \ep}
+ \left(\api\right)^2\,\left[{\beta_0^2\over \ep^2} 
  - {\beta_1\over \ep}\right] + \order{\alpha_s^3}\,.
\end{split}
\end{equation}
Here, $\beta(\alpha_s)$ is the \qcd\ beta function:
\begin{equation}
\begin{split}
\beta(\alpha_s) &= -\api\beta_0 - \left(\api\right)^2\beta_1 +
\order{\alpha_s^3}\,,\\
\beta_0 &= {1\over 4}\left(11 - {2\over 3} n_f\right)\,,\qquad
\beta_1 = {1\over 16}\left(102 - {38\over 3} n_f\right)\,.
\end{split}
\end{equation}

In the following sections, we will list the results for $\rho_A$,
$\rho_B$, and $\rho_C$ in terms of the bare coupling $\alpha_s^\bare$
which is related to the $\msbar$ renormalized coupling by
\begin{equation}
\begin{split}
\alpha_s^\bare &= {\cal N}^{-1} Z_\alpha(\alpha_s)\,\alpha_s\,,\\
\mbox{where}\qquad
{\cal N} &\equiv \exp[\ep(-\gamma_{\rm E} + \ln 4\pi)]\,,
\end{split}
\end{equation}
and $\gamma_{\rm E} = 0.577216\ldots$ is the Euler constant.  The charge
renormalization constant $Z_\alpha(\alpha_s)$ is
\begin{equation}
\begin{split}
Z_\alpha(\alpha_s) &= 1 - \api\,{\beta_0\over\ep} 
+ \left(\api\right)^2\,\left({\beta_0^2\over \ep^2} 
  - {\beta_1\over 2\ep}\right) + \order{\alpha_s^3}\,.
\end{split}
\end{equation}
%


\subsection{Two-loop virtual corrections}
The two-loop virtual corrections have been calculated in
\cite{harlander}. Typical diagrams are shown in \fig{fig::twoloop}.
Since gluons are massless, these diagrams depend on only a single scale
$q^2 = M_H^2$, where $q$ is the momentum of the Higgs boson.  They were
calculated using the method of ref.~\cite{baismi} which allows one to 
relate the integration-by-parts identities for the two-loop three-point
functions of~\fig{fig::twoloop} to those for three-loop two-point
functions~\cite{IP}.  The coding of the integration routines was done in
{\code FORM}~\cite{form} and was based on the program {\code
  MINCER}~\cite{mincer}.  It is worth mentioning that the resulting
program is capable of dealing with much more general diagrams than the
ones considered for this calculation. In particular it solves integrals
with (in principle) arbitrary integer powers of the propagators. This
may become useful in future applications where expansions have to be
applied to the diagrams.

The expression for $\rho_A$ is given in ref.~\cite{harlander}:
\begin{equation}
  \begin{split}
    \rho_A &= \rho_A^{(0)} + \apib\,\rho_A^{(1)} +
    \left(\apib\right)^2\,\rho_A^{(2)} + \order{\alpha_s^3}\,,
    \label{virt1}
  \end{split}
\end{equation}
\begin{equation}
  \begin{split}
    \rho_A^{(0)}(x) &= (1-\ep)\,\delta(1-x)\,,\\
    \rho_A^{(1)}(x) &= {\cal N}\left({\mu^2\over M_H^2}
        \right)^\ep\,\delta(1-x)\,\left[ -{3\over\ep^2} +
      {3\over\ep} + {21\over 2}\zeta_2
      + \ep\left(- 3 - {21\over 2}\zeta_2 + 7\zeta_3\right)\right]\,,\\
    \rho_A^{(2)}(x) &= {\cal N}^2\left({\mu^2\over M_H^2}
        \right)^{2\ep}\,\delta(1-x)\, \bigg\{{9\over2\ep^4}
    +{1\over\ep^3}\left(-{105\over 16}+{1\over 8}n_f\right) \\
    & \quad +{1\over\ep^2}\left(-{17\over8}-{243\over8}\zeta_2
      +{1\over12}n_f\right) \\
    &\quad +{1\over\ep}\bigg[{553\over48}+{981\over16}\zeta_2
    -{159\over8}\zeta_3+n_f\left(-{53\over72}-{15\over8}\zeta_2\right)\bigg]\\
    &\quad +{6749\over144}+{255\over8}\zeta_2+{85\over4}\zeta_3
    +{1161\over20}\zeta_2^2 +n_f\left(-{137\over27}
      -{5\over4}\zeta_2-{7\over4}\zeta_3\right)\bigg\}\,,
    \label{virt2} 
  \end{split}
\end{equation}
where $\zeta_n \equiv \zeta(n)$ is Riemann's $\zeta$ function which
takes the particular values
\begin{equation}
\begin{split}
 \zeta_2 &= \pi^2/6 = 1.64493\ldots\\
 \zeta_3 &= 1.20206\ldots.
\end{split}
\end{equation}


\subsection{Single real radiation}
\label{sec::singlereal}
The one-loop amplitudes for the processes ($ii$) in
\sct{sec::subprocesses} were computed in ref.~\cite{schmidt} in the form of
helicity amplitudes to order $\ep^0$.  Unfortunately, these amplitudes
are not sufficient for our purpose.  The integration over two-particle
phase space generates infrared singularities which take the form of
single and double poles in $\ep$ which combine with the $\order{\ep}$
and $\order{\ep^2}$ pieces of the amplitude to produce finite
contributions to the cross section.  If we were to work in the four
dimensional helicity scheme~\cite{4DH}, where gluons have two
polarization states, helicity amplitudes extended to higher order in
$\ep$ would suffice.  However, since we work in the conventional
dimensional regularization scheme, gluons have $d-2=2-2\ep$
polarizations.  Thus, in addition to helicity amplitudes we also need
amplitudes corresponding to the extra ``$\ep$'' polarizations.  Both the
additional pieces of the amplitudes and the integration over the two-particle
phase space have been evaluated
and will be presented elsewhere \cite{kilgore}. That
calculation includes the full dependence on the center-of-mass energy.
Here we are interested in the soft limit, to which we find that only the
$gg\to Hg$ process contributes.  The result is
\begin{equation}
  \begin{split}
    \rho_B &= \apib\,\rho_B^{(1)} + \left(\apib\right)^2\,\rho_B^{(2)}
    + \order{\alpha_s^3}\,,\\
  \end{split}
\end{equation}
\begin{equation}
  \begin{split}
  \label{eqn::tauB}
    \rho_B^{(1)}(x) &= {\cal N}\left({\mu^2\over M_H^2}\right)^\ep
    f_{2\ep}(x)\,\left[
      {3\over\ep^2}-{3\over\ep}-{9\over2}\zeta_2
      + \ep\left({9\over2}\zeta_2-7\zeta_3\right)\right]\,,\\
    \rho_B^{(2)}(x) &=
    {\cal N}^2\left({\mu^2\over M_H^2}\right)^{2\ep} 
    \bigg\{
    f_{2\ep}(x)\,
    \bigg[ -{9\over\ep^4}+{9\over\ep^3}
    +{45\over\ep^2}\,\zeta_2 \\
    &\quad\quad
    +{1\over\ep}\left( - 9 - 45\,\zeta_2 + 42\,\zeta_3\right)
    - 18 - 42\,\zeta_3 - {603\over10}\,\zeta_2^2\bigg]\\
    &\quad + f_{4\ep}(x)
    \left[-{9\over 8\ep^4} + {9\over 8\ep^3}
      + {63\over 8\ep^2}\,\zeta_2
      + {1\over\ep} \left( -{63\over8}\,\zeta_2 + 21\,\zeta_3\right)
      -21\,\zeta_3 + {189\over 80} \,\zeta_2^2\right]
    \bigg\}\,.
  \end{split}
\end{equation}
Expanding $f_{a\ep}(x)$ (see \eqn{eqn::distrdef}), \eqn{eqn::tauB}
describes both delta function and large logarithm contributions to
the radiative corrections.


\subsection{Double real radiation}
\label{sec::doublereal}
The tree-level amplitudes for the processes ($iii$) of
\sct{sec::subprocesses} with two partons in the final state was
evaluated in \cite{dawkau}, though again only as helicity amplitudes to
$\order{\ep^0}$.  We have evaluated the full $\ep$ dependence in the 
conventional dimensional regularization scheme by computing all of the
Feynman diagrams, squaring the amplitude and integrating over phase
space, dropping terms that do not 
contribute in the soft limit as described in \sct{sec::softapprox}.
We find that only the $gg\to Hgg$ and $gg\to Hq\bar{q}$ processes
contribute with the result
\begin{equation}
  \begin{split}
    \rho_C(x) &=
    {\cal N}^2\left(\apib\right)^2
    \left({\mu^2\over M_H^2}\right)^{2\ep} f_{4\ep}(x)
    \bigg\{
    {45\over 8\ep^4} - {57\over 16\ep^3} + {1\over\ep^2}\left(
      {17\over8} - {81\over2}\zeta_2\right) \\
    &\quad + {1\over\ep}\left({203\over48}
      + {417\over16}\zeta_2 - {975\over8}\zeta_3\right)
    + {76\over9} - {119\over8}\zeta_2 + {317\over4}\zeta_3
    - {333\over80}\zeta_2^2 \\
    &\quad + n_f\bigg[
    - {1\over8\ep^3} - {1\over12\ep^2}
    + {1\over\ep}\left(-{13\over72} + {7\over8}\zeta_2\right)
    - {10\over27} + {7\over12}\zeta_2
    + {31\over12}\zeta_3\bigg]
\bigg\} + \order{\alpha_s^3}\,.
  \end{split}
\end{equation}
The terms proportional to $n_f$ come from the $gg\to Hq\bar q$
sub-process.  (See \fig{fig::diagrams}.)


\subsection{Mass factorization}
After combining all processes and renormalizing, there are still
infrared singularities left over.  These are all associated with mass
factorization and can be absorbed into process independent functions
$\Gamma_{ij}(x)$ associated with the incoming partons.  The
({\abbrev IR}-finite) partonic cross section,
$\hat{\sigma}$, is defined implicitly by the relation
\begin{equation}
\begin{split}
\label{eqn::massfact}
  \sigma_{ij} &= \sum_{\bar\imath,\bar\jmath}
  \hat{\sigma}_{\bar\imath\bar\jmath}\otimes
  \Gamma_{\bar\imath{i}}\otimes\Gamma_{\bar\jmath{j}},
\end{split}
\end{equation}
where $\sigma$ is defined in \eqn{eq::sigrho}.
The symbol $\otimes$ indicates convolution over longitudinal
momentum fractions,
\begin{equation}
\begin{split}
\Lx f\otimes g\Rx(x) &= \int_0^1 dy\int_0^1 dz\,f(y)\,g(z)\,\delta(x-yz),
\end{split}
\end{equation}
and the subscripts $\bar\imath$, $i$, etc., indicate parton identities.
The $\Gamma_{ij}(x)$'s are given by
\begin{equation}
\begin{split}
\Gamma_{ij}(x) &= \delta_{ij}\delta(1-x) - 
   {\alpha_s\over\pi}{P^{(0)}_{ij}(x)\over\ep}\\
   &\quad + \Lx{\alpha_s\over\pi}\Rx^2\LB
     {1\over2\ep^2}\Lx\Lx P^{(0)}_{ik}\otimes P^{(0)}_{kj}\Rx(x)
     + \beta_0P^{(0)}_{ij}(x)\Rx - {1\over2\ep}P^{(1)}_{ij}(x)\RB
      +{\cal O}(\alpha_s^3),
\end{split}
\end{equation}
where the $P_{ij}^{(n)}$ are the $n$th order splitting functions
\cite{splitting,EllVog}. In the soft limit, only $P^{(n)}_{ij}$ with
$i=j=g$ contribute. They are given by
\begin{equation}
\begin{split}
  P^{(0)}_{gg}(x)\issoft\ & \left({11\over 4} 
    - {n_f\over 6}\right)\delta(1-x) + 3\,{\cal D}_0(x)\,,\\
  P^{(1)}_{gg}(x)\issoft\ & 
  \left( 6 + {27\over 4}\zeta_3 - {2\over 3}n_f\right)\,\delta(1-x) 
  + \left({67\over 4} - {9\over 2}\,\zeta_2 
    - {5\over 6}\,n_f\right)\,{\cal D}_0(x)\,.
\end{split}
\end{equation}

Writing
\begin{equation}
\begin{split}
\sigma_{gg} = \sigma_{gg}^{(0)} + \api\sigma_{gg}^{(1)} +
\left(\api\right)^2\sigma_{gg}^{(2)} + \ldots\,,
\end{split}
\end{equation}
and similarly for $\hat\sigma_{gg}$, \eqn{eqn::massfact} can be solved
for $\hat\sigma_{gg}$ order by order to give
\begin{equation}
\begin{split}
  \hat\sigma_{gg}^{(0)}(x) &= \sigma_{gg}^{(0)}(x) \equiv \sigma_0\,\delta(1-x)\,,\\
  \hat\sigma_{gg}^{(1)}(x) &= \sigma_{gg}^{(1)}(x)
  + 2\,\sigma_0\, P^{(0)}_{gg}(x)\, {1\over \ep}\,, \\
  \hat\sigma_{gg}^{(2)}(x) &= \sigma_{gg}^{(2)}(x)
    - {1\over \ep^2}\sigma_0\left( \beta_0\, P^{(0)}_{gg}(x) 
      + 2\,(P_{gg}^{(0)}\otimes P_{gg}^{(0)})(x) \right) \\
  &\quad\quad    + {1\over \ep}\left( \sigma_0 P_{gg}^{(1)}(x) 
      + 2\,(P_{gg}^{(0)}\otimes \hat\sigma_{gg}^{(1)})(x) 
      \right)\,,
\label{eq::sigP}
\end{split}
\end{equation}
where
\begin{equation}
\begin{split}
  \sigma_0 \equiv {\pi\over 576 v^2}\,\left(\api\right)^2\,(1 + \ep +
  \ep^2)\,.
\end{split}
\end{equation}
The convolutions
in \eqn{eq::sigP} can be evaluated with the help of the following relations
\begin{equation}
\begin{split}
  ({\cal D}_0\otimes{\cal D}_0)(x) \issoft&\ -\zeta_2\,\delta(1-x) 
  + 2\,{\cal D}_1(x)\,,\\
  ({\cal D}_0\otimes{\cal D}_1)(x) \issoft&\ \zeta_3\,\delta(1-x)
  - \zeta_2\,{\cal D}_0(x) + {3\over 2}{\cal D}_2(x)\,,\\
  ({\cal D}_0\otimes{\cal D}_2)(x) \issoft&\
  -{4\over 5}\,\zeta_2^2\,\delta(1-x) + 2\,\zeta_3\,{\cal D}_0(x)
  - 2\,\zeta_2\,{\cal D}_1(x) + {4\over 3}{\cal D}_3(x)\,,\\
  ({\cal D}_1\otimes{\cal D}_1)(x) \issoft&\
  -{1\over 10}\zeta_2^2\,\delta(1-x) - 2\zeta_2{\cal D}_1(x) + {\cal
    D}_3(x)\,.
\end{split}
\end{equation}


\subsection{Partonic cross section}
Putting things together, we finally arrive at
\begin{equation}
\begin{split}
    \hat\sigma^{(0)}_{gg} &= {\pi\over 576v^2}\,\left(\api\right)^2
       \delta(1-x)\,,\\
    \hat\sigma^{(1),{\rm soft}}_{gg} &= {\pi\over 576v^2}
       \left(\api\right)^2\left\{
      \delta(1-x)\LB {11\over2} + 6\,\zeta_2\RB
       - 6\,{\cal D}_0(x)\,\logmuh{}
       + 12\,{\cal D}_1(x)\right\}\,,\\
    \hat\sigma^{(2),{\rm soft}}_{gg} &= {\pi \over 576 v^2}
       \left(\api\right)^2\bigg\{
       \delta(1-x)\LB{11399\over144} + {133\over2}\zeta_2
          - {165\over4}\zeta_3 - {9\over20}\zeta_2^2\right.\\
       & \quad\mbox{} + \Lx{27\over2} + {33\over2}\zeta_2
         - {171\over2}\zeta_3 \Rx\logmuh{} - 18\zeta_2\,\logmuh{2}
         +{19\over8}\logmut{} \\
       & \quad\mbox{}+ n_f\left(-{1189\over144} - {5\over3}\zeta_2
         + {5\over6}\zeta_3 + \Lx -{11\over6}
         - \zeta_2\Rx\logmuh{} + {2\over3}
         \logmut{}\right)\bigg]\\
       &+{\cal D}_0(x)\bigg[
       -{101\over3} + 33\,\zeta_2 +{351\over2}\zeta_3
        +\left(- {133\over2}+ 45\zeta_2\right)\logmuh{}
        - {33\over4}\logmuh{2} \\
       &\quad +\mbox{} n_f \left({14\over9} - 2\,\zeta_2
        + {5\over3}\logmuh{} + {1\over2}\logmuh{2}\right)\bigg]\\
       &+{\cal D}_1(x)\LB 133 - 90\,\zeta_2 + 33\,\logmuh{}
           + 36\,\logmuh{2}
           + n_f\left( - {10\over3} - 2\,\logmuh{} \right)
           \RB\\
       &+{\cal D}_2(x)\LB-33 - 108\,\logmuh{} +2n_f \RB
        + 72\,{\cal D}_3(x)\bigg\}\,,
\label{eq::sigmahat}
\end{split}
\end{equation}
with $\logmuh{}\equiv\ln(\mu^2/M_H^2)$ and $\logmut{}\equiv
\ln(\mu^2/M_t)$.  As a numerical example, let us set $\mu^2=M_H^2$ and
insert the values $M_H = 130$\,GeV, $M_t=175$\,GeV, and $\alpha_s =
0.112$:
\begin{equation}
\begin{split}
\hat\sigma_{gg} = {\pi \over 576 v^2}\left(\api\right)^2&\bigg\{
[1 + 0.548 + 0.107 ]\, \delta(1-x)
+ [0 + 0 + 0.283]\,{\cal D}_0(x) \\&
+ [0 + 0.428 - 0.040]\,{\cal D}_1(x)
+ [0 + 0 - 0.029]\,{\cal D}_2(x) \\&
+ [0 + 0 + 0.092]\,{\cal D}_3(x)
\bigg\} + {\cal O}(\alpha_s^5)\,,
\end{split}
\end{equation}
where the individual numbers in square brackets correspond to
the $\hat\sigma_{gg}^{(0)}$, $\hat\sigma_{gg}^{(1)}$, and
$\hat\sigma_{gg}^{(2)}$ contributions. 
We observe that the coefficient of the $\delta$ function exhibits
satisfactory convergence behavior. The magnitude of its \nnlo\
contribution agrees roughly with the estimate of ref.~\cite{harlander}.

\subsubsection{Checks on the result}
There are several checks that can be performed on our result.  The first
check is to observe that all poles in $\ep$ cancel to give us a finite
result for $\hat\sigma^{(2),{\rm soft}}_{gg}$.  Since the ${\cal
  D}_n(x)$ terms at order $\ep^0$ are linked to the poles of the real
emission contributions, this already gives us great confidence in these
terms. Below we will discuss an even more stringent constraint on the ${\cal
  D}_n$ coefficients.

Another check is to compare the three leading poles of the virtual
two-loop corrections to the general result of ref.\,\cite{catani}, where
they are expressed in terms of universal functions that depend only on
the identity of the external partons. The poles of the diagrammatic
result of~\cite{harlander} fully obey this observation~\cite{hkdpf2000}.

One can now combine the universal (three leading) \nnlo\ pole terms
with the \nlo\ results $\rho_A^{(0,1)}$ and $\rho_B^{(1)}$ to derive
the coefficients of ${\cal D}_n(x)$ for $n=1,2,3$ in
$\hat\sigma_{gg}^{(2),\rm soft}$.  Note that this uses only one-loop
results and the universal behavior of \qcd\ amplitudes and is done
without reference to any of the newly calculated \nnlo\ terms
(including those of \cite{harlander}). The pole structure of the real
radiation terms must take the form 
\begin{equation}
\begin{split}
  \left(\rho_B^{(2)} + \rho_C^{(2)}\right)_{\rm pole} &=
  f_{2\ep}(x)\left({\mu^2\over M_H^2}\right)^{2\ep}
    \left[ {a_{24}\over \ep^4} + {a_{23}\over \ep^3} +
    {a_{22}\over \ep^2} + {a_{21}\over \ep} \right]\\
  &\quad +f_{4\ep}(x)\left({\mu^2\over M_H^2}\right)^{2\ep}
    \left[ {a_{44}\over \ep^4} +
    {a_{43}\over \ep^3} + {a_{42}\over \ep^2} + {a_{41}\over \ep}
  \right]\,.
\label{eq::deltarad}
\end{split}
\end{equation}
Expanding the $f_{n\ep}(x)$ terms as in \eqn{eqn::distrdef} and
requiring that all poles vanish from $\hat\sigma^{(2)}_{gg}$, we can
solve for $a_{n4}$, $a_{n3}$ and $a_{n2}$ ($n = 2,4$), while $a_{21}$
and $a_{41}$ remain undetermined.  This solution is sufficient to fix
the coefficients of ${\cal D}_1(x)$, ${\cal D}_2(x)$ and ${\cal D}_3(x)$
in $\hat\sigma^{(2)}_{gg}$, though not the coefficient of ${\cal D}_0(x)$.
We find complete agreement between the ${\cal D}_{\{3,2,1\}}$ terms
determined in this way and those found in \eqn{eq::sigmahat}.

Let us note that the absence of terms $\propto f_{n\ep}$ with $n\neq 2,
4$ in~\eqn{eq::deltarad} can be understood from the following
consideration. The index $n$ of the $f_{n\ep}$ terms has two sources:
the phase space 
elements for single and double real radiation contribute factors of
$(1-x)^{1-2\ep}$ and $(1-x)^{3-4\ep}$ respectively (see {\it e.g.}
Appendix E of ref.~\cite{DYsoft}), while the one-loop corrections to
single real radiation amplitudes squared can contribute factors of
$(1-x)^{m}$, $(1-x)^{m-\ep}$ or $(1-x)^{m-2\ep}$, where $m$ is an
integer~\cite{schmidt,kilgore}.  [Tree-level amplitudes squared contain
only integer powers of $(1-x)$.]  The $(1-x)^{-2-\ep}$ terms in the
one-loop amplitude squared should generate an $f_{3\ep}(x)$
contribution. However, the universal functions describing the soft
factorization properties of one-loop amplitudes only contain terms that
scale like $(1-x)^{-1}$ and $(1-x)^{-1-2\ep}$~\cite{SoftFact}.  Thus,
there are no terms in the squared amplitude that scale like
$(1-x)^{-2-\ep}$ and that survive the soft limit.

Another way to derive the ${\cal D}_{\{1,2,3\}}(x)$ terms without using
any process-specific \nnlo\ results is to perform a threshold
resummation analysis along the lines of \cite{sglue,gghresum}.  Again we find
complete agreement with our result.  As a by-product, we can use our
result for the coefficient of ${\cal D}_0$ in order to derive additional
input for the resummation formula.  Of particular interest is the \nnlo\ 
coefficient attributed to large-angle soft-gluon emissions which has
been derived for the Drell-Yan and deep inelastic scattering processes
in ref.\,\cite{Vogt} (where it is called $D_2^P$, $P={\rm DY},{\rm
  DIS},\ldots$). Following this analysis for $gg\to H$, one finds that
the corresponding coefficient can be obtained from the one in the
Drell-Yan process [Eq.\,(21) in \cite{Vogt}] simply by changing the
color factor $C_F$ to $C_A$.\footnote{We thank W.~Vogelsang for pointing
  this out to us.}

A further check concerns the terms proportional to $\ln\mu^2$ which can
be derived from renormalization group and factorization scale invariance
of the physical cross section:\footnote{Note that we restrict ourselves
  to the limit of soft gluons.}
\begin{equation}
\begin{split}
\sigma_{pp} = \hat\sigma_{gg}\otimes g \otimes g\,,
\end{split}
\end{equation}
where $g$ is the gluon distribution function.  This leads to
\begin{equation}
\begin{split}
  \mu^2\ddoverdd{\mu^2}\,\hat\sigma_{gg} + 2\left[\api P^{(0)}_{gg} +
    \left(\api\right)^2P_{gg}^{(1)}\right]
  \otimes\hat\sigma_{gg} = \order{\alpha_s^5}\,.
\end{split}
\end{equation}
\Eqn{eq::sigmahat} satisfies this relation.
%



\section{Discussion}
For a consistent discussion of the physical implications at \nnlo, one
has to convolute the partonic cross section at a given order with the
corresponding parton distribution functions (\pdf{}s) of the same
order. Unfortunately, {\abbrev NNLO} \pdf{}s are not yet
available.\footnote{Various necessary ingredients are already
available, and it seems to be clear that the combined efforts of
different groups will eventually result in an appropriate set of
distributions.}
Thus, we will restrict ourselves to a semi-quantitative investigation
of the {\abbrev NNLO} effects by using {\abbrev NLO} \pdf{}s at \nnlo.
In all results we have used the {\abbrev CTEQ5}~\cite{cteq5} 
family of parton distributions.

We define
\begin{equation}
\begin{split}
  K_{\rm NLO} = {\sigma_{\rm NLO}\over \sigma_{\rm LO}}\,,\qquad
  K_{\rm NNLO} = {\tilde\sigma_{\rm NNLO}\over \sigma_{\rm LO}}\,,
  \label{eq::kfactor}
\end{split}
\end{equation}
where $\tilde\sigma_{\rm NNLO}$ is obtained by convoluting the partonic
$\hat\sigma^{(2)}_{gg}$ with {\abbrev NLO} \pdf{}s and using {\abbrev
  NLO} evolution of the strong coupling constant. $\sigma_{\rm LO}$ and
$\sigma_{\rm NLO}$ are evaluated consistently with {\abbrev LO} and
{\abbrev NLO} \pdf{}s and $\alpha_s$ evolution, respectively.  Of course
the treatment for $K_{\rm NNLO}$ is inconsistent, but one may
nevertheless get an idea on the magnitude of the effects.


%
\begin{figure}
  \begin{center}
    \leavevmode
    \begin{tabular}{cc}
      \epsfxsize=7.cm
      \epsffile[110 265 465 560]{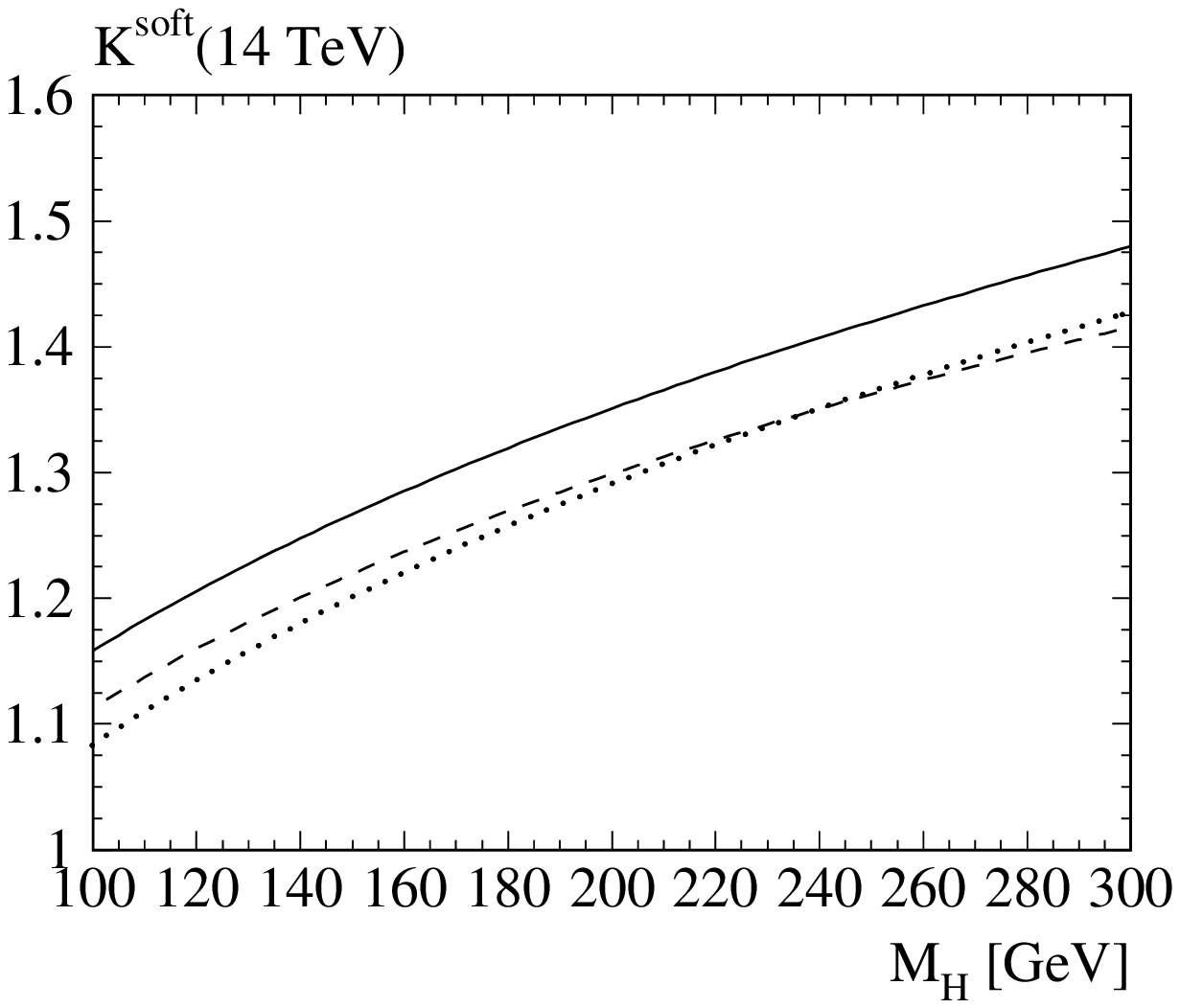} &
      \epsfxsize=7.cm
      \epsffile[110 265 465 560]{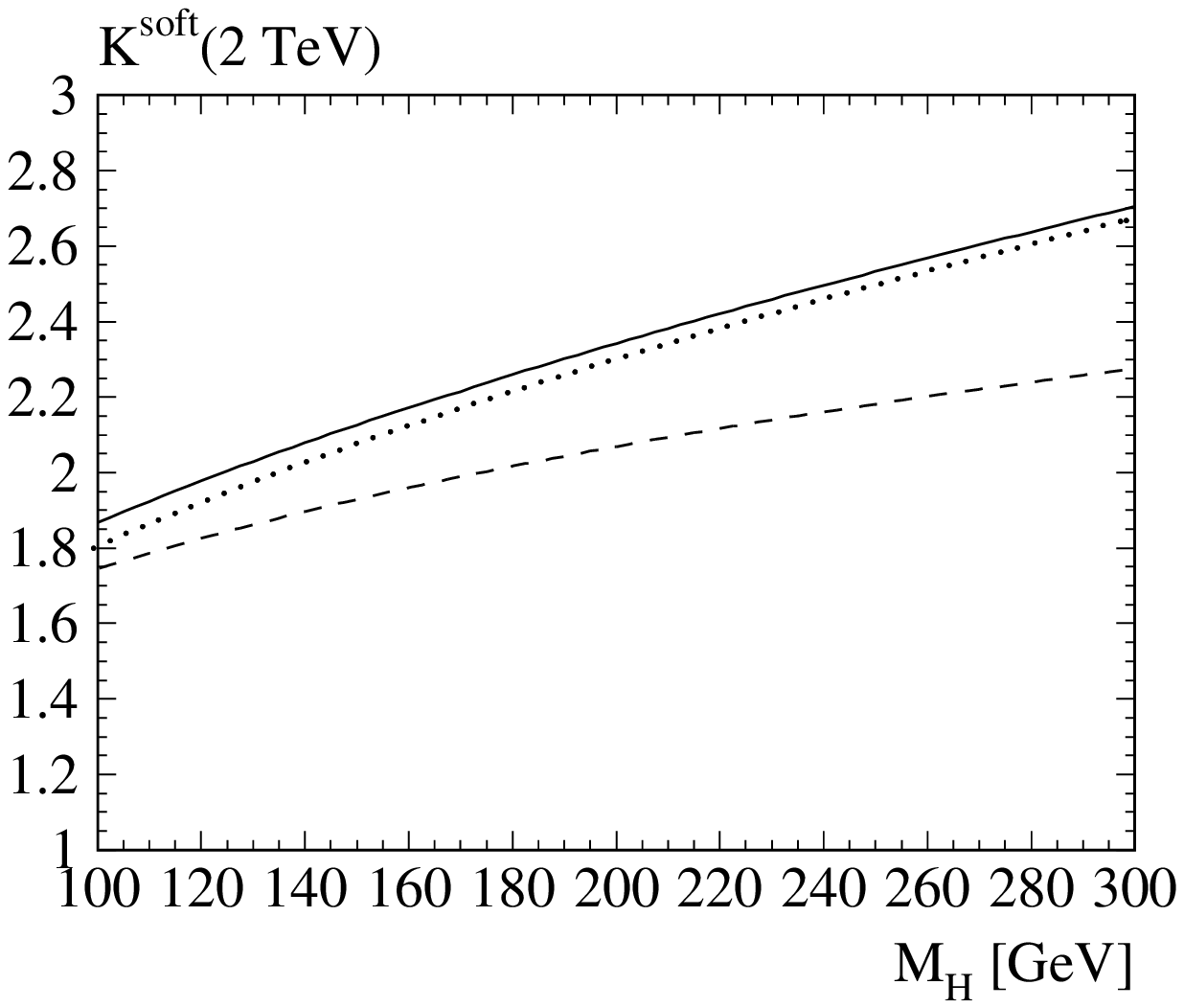}\\
      (a) & (b)
    \end{tabular}
    \parbox{14.cm}{
      \caption[]{\label{fig::khksoft}\sloppy
        $K$ factor as defined in \eqn{eq::kfactor}, using the purely
        soft approximation of \eqn{eq::sigmahat}.  Dashed and solid
        lines correspond to {\abbrev NLO} and {\abbrev NNLO},
        respectively. The dotted line represents the approximate result
        $\bar\sigma_{gg}^{\rm soft}$ of \cite{gghresum}.\\
        (a): $\sqrt{S} = 14$\,TeV, (b): $\sqrt{S} = 2$\,TeV, where
        $\sqrt{S}$ is the c.m.s. energy of the proton-proton
        system.}}
  \end{center}
\end{figure}
%


%
\begin{figure}
  \begin{center}
    \leavevmode
    \begin{tabular}{cc}
      \epsfxsize=7.cm
      \epsffile[110 265 465 560]{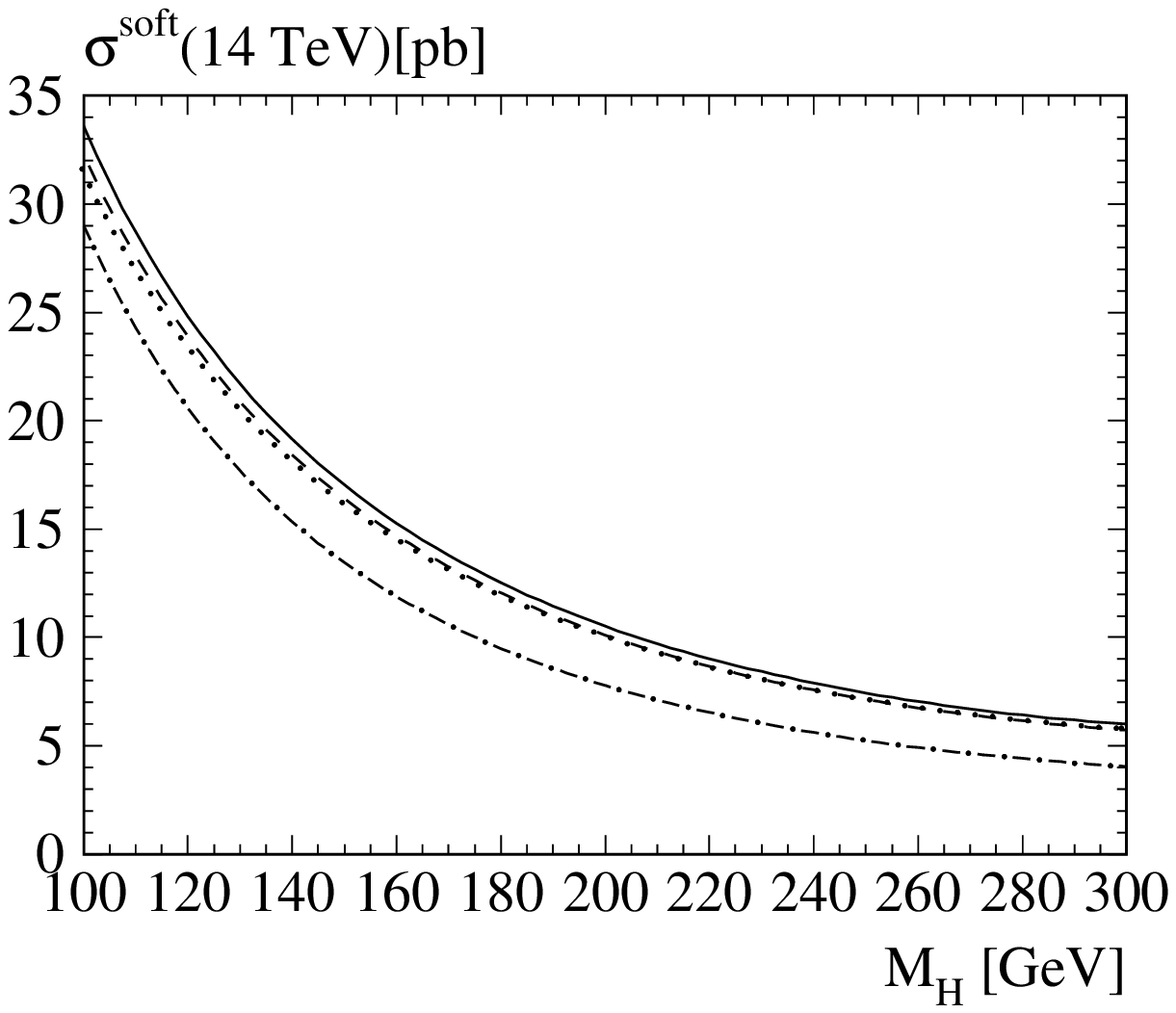} &
      \epsfxsize=7.cm
      \epsffile[110 265 465 560]{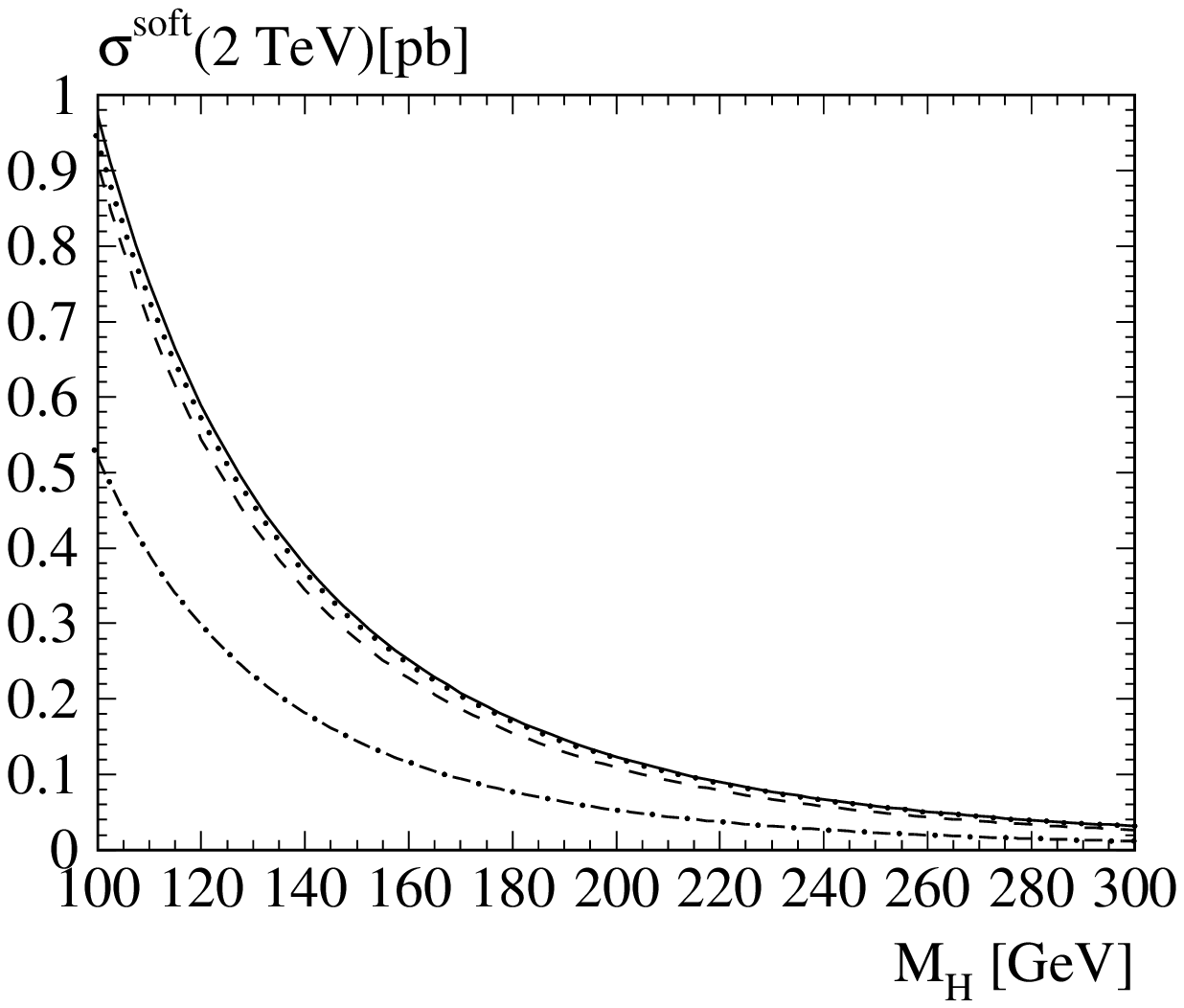}
    \end{tabular}
    \parbox{14.cm}{
      \caption[]{\label{fig::sigmasoft}\sloppy
        Cross section $\sigma(pp\to H+X)$ in the purely soft
        approximation ({\it cf.\/} caption of \fig{fig::khksoft}).
        Dash-dotted, dashed, and solid line correspond to {\abbrev LO},
        {\abbrev NLO}, and {\abbrev NNLO} results obtained from
        \eqn{eq::sigmahat}.  The dotted line is the soft part of the
        {\abbrev NNLO} approximation of ref.~ \cite{gghresum}.  (a):
        $\sqrt{S} = 14$\,TeV, (b): $\sqrt{S} = 2$\,TeV.  }}
  \end{center}
\end{figure}
%


\Fig{fig::khksoft} shows the $K$-factors $K_{\rm NLO}$ and $K_{\rm
  NNLO}$ for a proton-proton c.m.s. energy of (a)~14~TeV and
(b)~2~TeV, when only the purely soft terms [{\it i.e.\/} $\propto
  \delta(1-x)$ 
and $\propto {\cal D}_n(x)$] are kept.  One finds a nicely convergent
behavior when going from {\abbrev NLO} to {\abbrev NNLO}, both for
{\abbrev LHC} and Tevatron energies.  Also shown are the
approximate soft {\abbrev NNLO} terms of \cite{gghresum} (see below).

\Fig{fig::sigmasoft} shows the cross section $\sigma(pp\to H+X)$
obtained by weighting the terms in \eqn{eq::sigmahat} by the ratio of
the \lo\ expression for $\hat\sigma$ including full top mass
dependence given in \eqn{eqn::lo} with the \lo\ result in the heavy
top limit.  This procedure has been shown to be an excellent 
approximation of the full top mass dependence at \nlo\ for Higgs
boson masses up to and even beyond the top threshold.  These curves again
demonstrate the nice convergence of the soft contributions to the
cross section.

Note that these numbers do not represent the
full phenomenological result, because of the missing, non-negligible
contributions coming from hard scattering.  Rather, this consistent
comparison of the soft limits at {\abbrev NLO} and {\abbrev NNLO}
should give an indication of the magnitude of the total corrections.

\subsection{Estimating the hard scattering contributions}
\begin{figure}
  \begin{center}
    \leavevmode
    \begin{tabular}{cc}
      \epsfxsize=7.cm
      \epsffile[110 265 465 560]{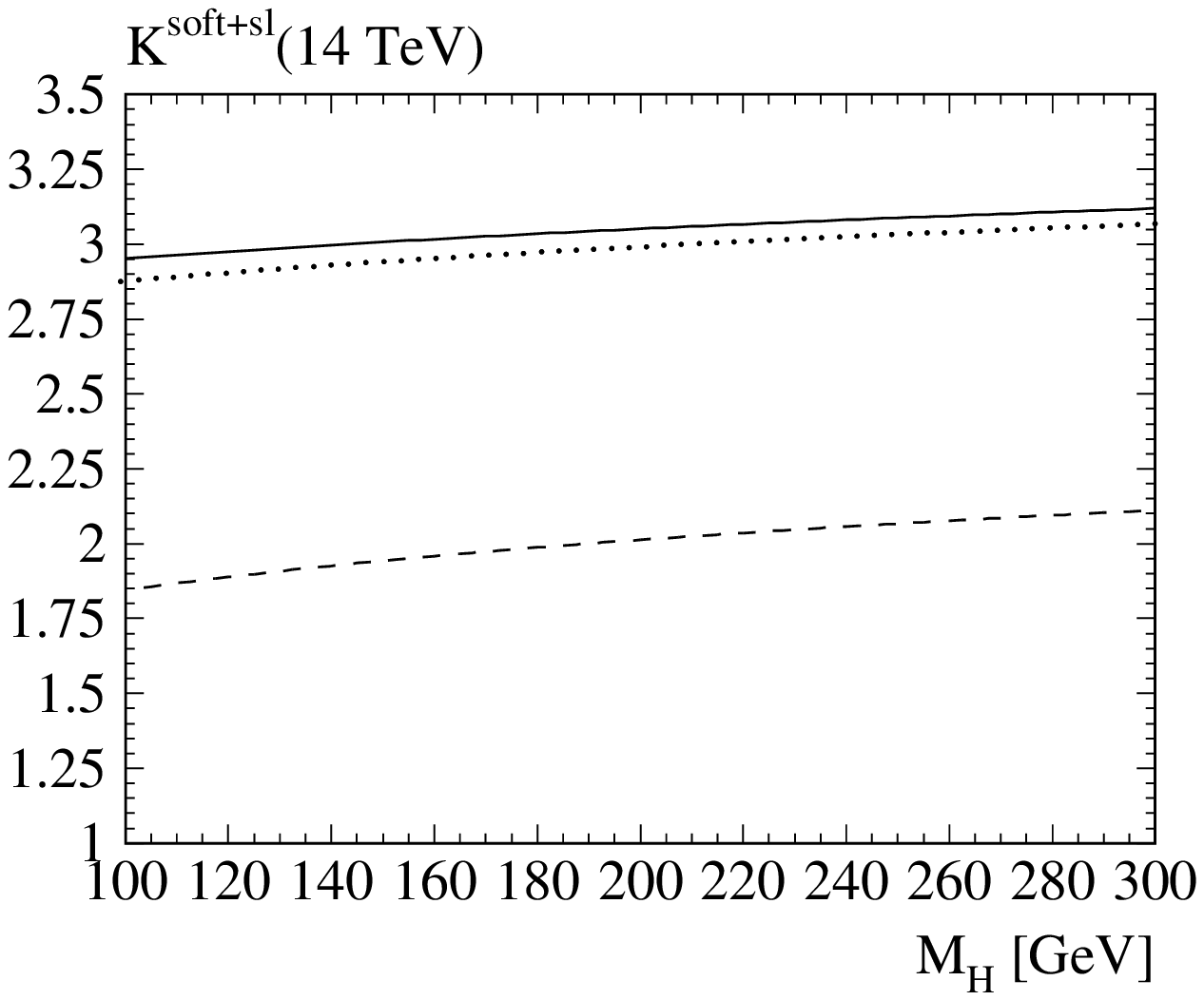} &
      \epsfxsize=7.cm
      \epsffile[110 265 465 560]{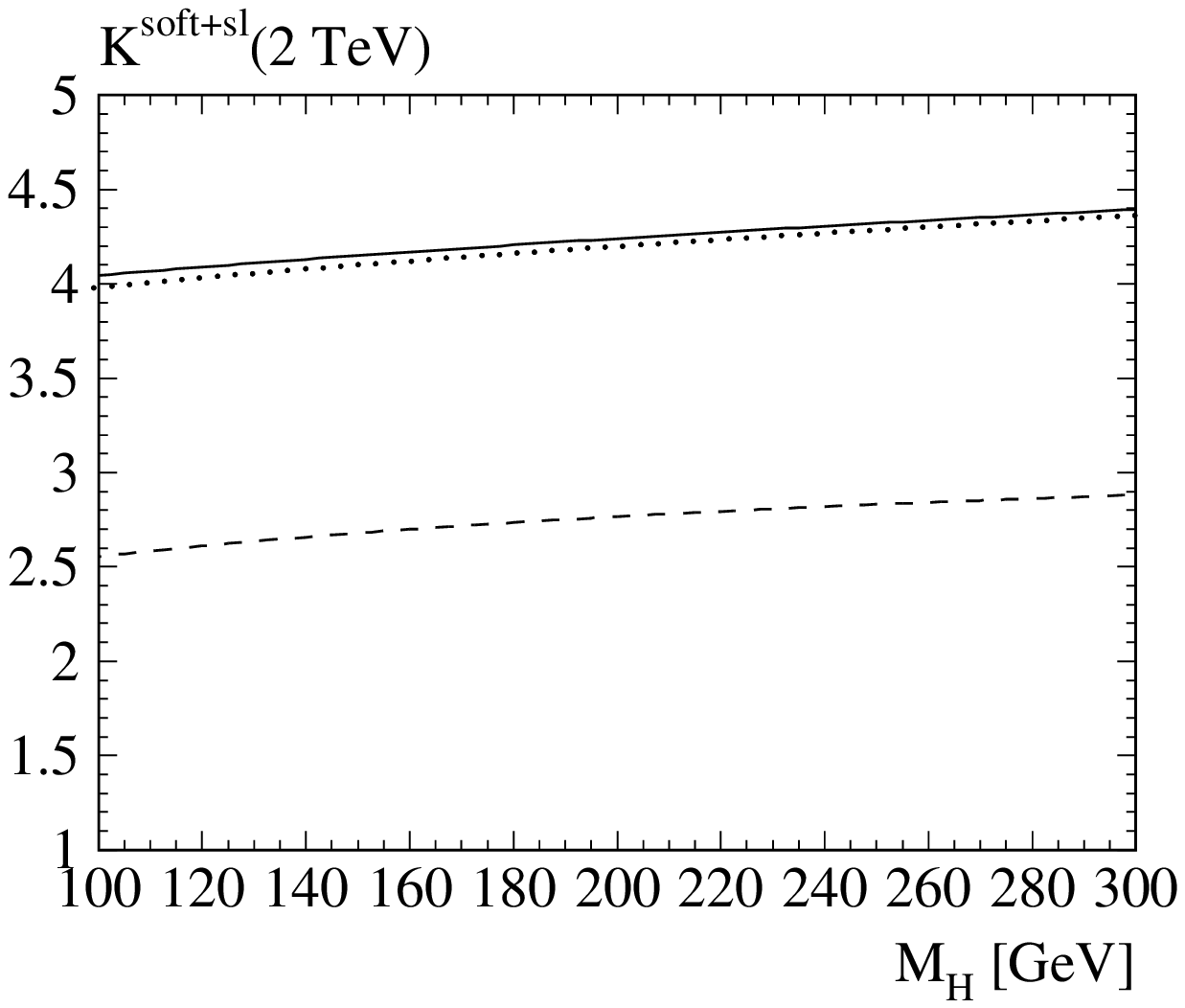}\\
      (a) & (b)
    \end{tabular}
    \parbox{14.cm}{
      \caption[]{\label{fig::khksoftsl}\sloppy
        $K$ factor as defined in \eqn{eq::kfactor}, including soft and
        ``sub-leading'' terms, {\it i.e.\/} using \eqn{eq::softsl} for the
        dashed and solid curves, and \eqn{eq::sigmabar} (with
        $\bar\sigma_{gg}^{\rm soft}$ extracted from \cite{gghresum})
        for the dotted one.}}
  \end{center}
\end{figure}

At \nlo\ one finds that the soft corrections, while significant, are
by no means dominant.  Comparison of the dashed lines in
\fig{fig::khksoft} to those of \fig{fig::khksoftsl} (discussed below),
shows that the soft terms account for no more than half (and often
significantly less) of the total \nlo\ correction.  Thus, it will be
necessary to compute the full \nnlo\ correction, including the hard
scattering terms, before one could conclude that the gluon fusion
process is under control.

Until such time as the full calculation is available, we rely on the
estimate of ref.~\cite{gghresum} which points out that the hard
scattering corrections at \nlo\ are dominated by formally sub-leading
terms of the form $\ln^i(1-x)$ and then estimates the \nnlo\
correction including these sub-leading terms.  We write their result
as 
\begin{equation}
\begin{split}
\bar\sigma_{gg} = \bar\sigma_{gg}^{\rm soft} + \bar\sigma_{gg}^{\rm sl}\,.
\label{eq::sigmabar}
\end{split}
\end{equation}
$\bar\sigma_{gg}$ is generated from an expansion of a one-loop
resummation formula and naturally the analytic form of
$\bar\sigma_{gg}^{\rm soft}$ differs from the actual {\abbrev NNLO}
result given in \eqn{eq::sigmahat}.  Similarly, the (formally
sub-leading, but numerically dominant) part $\bar\sigma_{gg}^{\rm sl}$
is only an 
approximation of the actual, as yet unknown expression.  It is
expected that the terms with the highest power of $\ln(1-x)$
[$\ln(1-x)$ at \nlo, $\ln^3(1-x)$ at \nnlo] obtained in this manner
are correct but that terms with lower powers of $\ln(1-x)$ are not.
In the case
of Drell-Yan production, the result of ref.~\cite{gghresum} {\it
including} the corresponding sub-leading terms $\bar\sigma_{q\bar
q}^{\rm sl}$ reproduces the full answer~\cite{DYfull} to within less
than $5$\%.  However, as can be seen in \fig{fig::khksoft}, the
{\it soft\/} terms generated for Higgs productions do not agree with
the true \nnlo\ soft terms as precisely, especially at \lhc\
energies.

With these caveats, we present an estimate of the full result by
combining our result for the soft correction with the approximate,
formally sub-leading logarithmic terms of ref.~\cite{gghresum}.  That
is, we are going to add $\bar\sigma_{gg}^{\rm sl}$ of
\eqn{eq::sigmabar} to our \eqn{eq::sigmahat} and write:
\begin{equation}
\begin{split}
\check\sigma_{gg}^{(i),{\rm soft}+{\rm sl}} = 
\hat\sigma_{gg}^{(i),{\rm soft}} + \bar\sigma_{gg}^{(i),{\rm sl}}\,,
\qquad i=1,2\,,
\label{eq::softsl}
\end{split}
\end{equation}
where~\cite{gghresum}
\begin{equation}
\begin{split}
\bar\sigma_{gg}^{(1),{\rm sl}} &= -24\,\ln(1-x)\,,\\
\bar\sigma_{gg}^{(2),{\rm sl}} &= 
-144\,\ln^3(1-x) + \ln^2(1-x)\,(138 - 216\,\logmuh{} - 4\,n_f) 
\\&\quad
+ \ln(1-x)\,(-132 + 138\,\logmuh{} - 72\,\logmuh{2} - 4\,\logmuh{}\,n_f 
+ 144\,\zeta_2)\,.
\label{eq::sl}
\end{split}
\end{equation}
Again, one must be very cautious about phenomenological interpretation
of this result, because of the aforementioned approximate nature of
$\bar\sigma_{gg}^{(2),{\rm sl}}$.


%
%


%
\begin{figure}
  \begin{center}
    \leavevmode
    \begin{tabular}{cc}
      \epsfxsize=7.cm
      \epsffile[110 265 465 560]{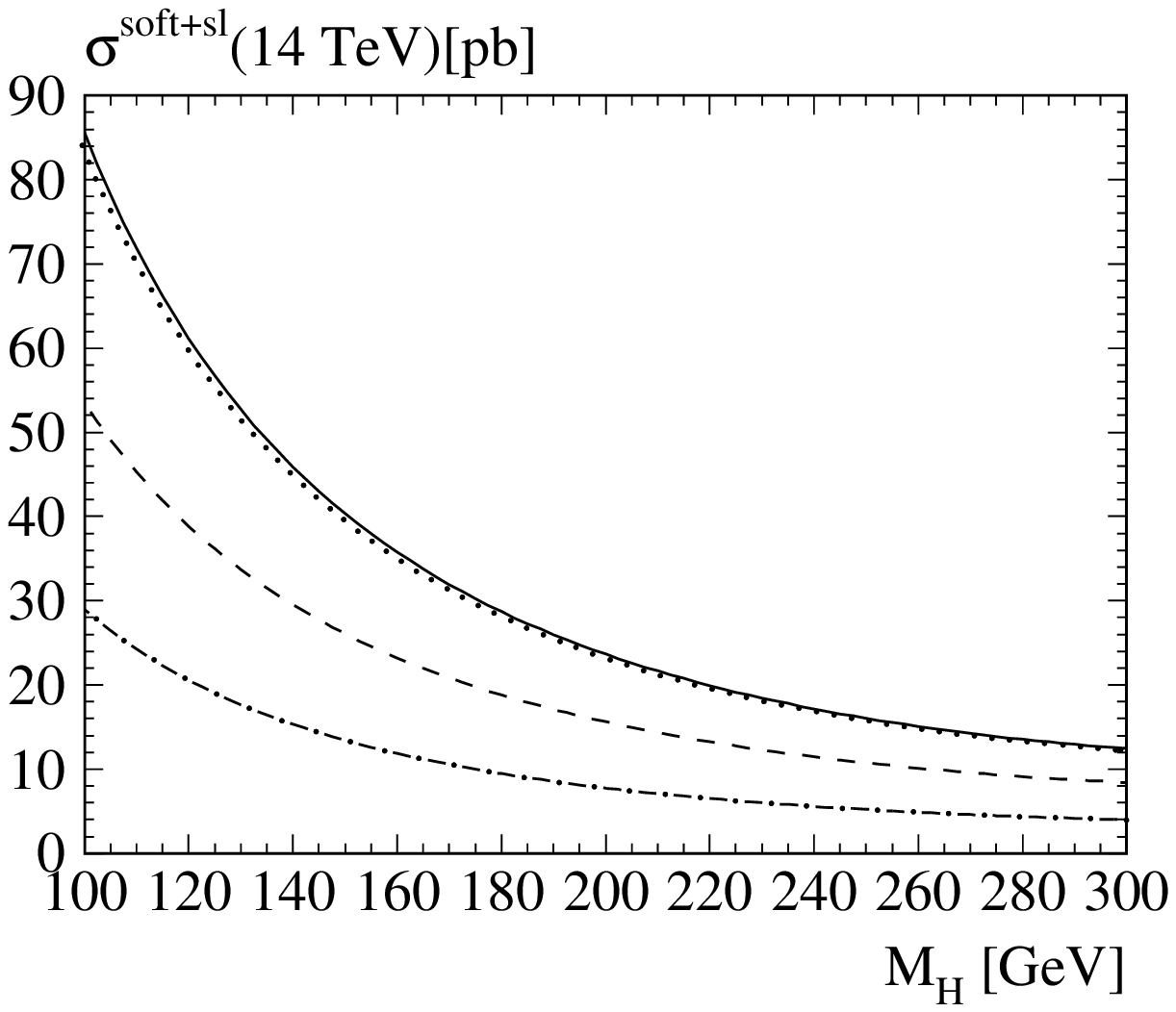} &
      \epsfxsize=7.cm
      \epsffile[110 265 465 560]{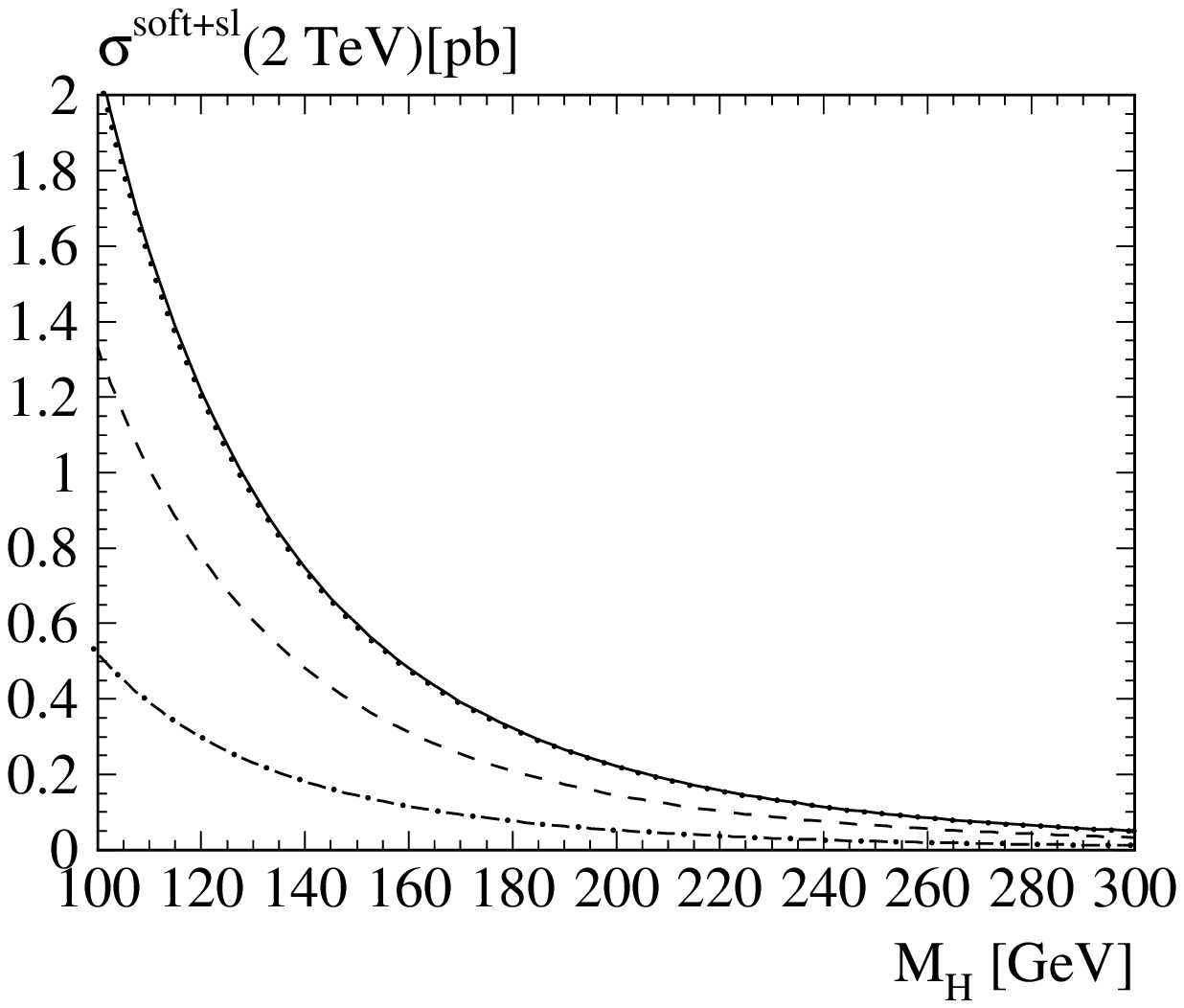}
    \end{tabular}
    \parbox{14.cm}{
      \caption[]{\label{fig::sigmasoftsl}\sloppy
        Cross section $\sigma(pp\to H+X)$ including soft and
        ``sub-leading'' terms ({\it cf.\/} caption of \fig{fig::khksoftsl}).
        Dash-dotted, dashed, and solid line correspond to {\abbrev LO},
        {\abbrev NLO}, and {\abbrev NNLO} results obtained from
        \eqn{eq::sigmahat}.  The dotted line is the {\abbrev NNLO}
        result based on the approximation given in \cite{gghresum}.
        (a): $\sqrt{S} = 14$\,TeV, (b): $\sqrt{S} = 2$\,TeV.  }}
  \end{center}
\end{figure}
%


In \fig{fig::khksoftsl} we show the corresponding $K$-factors $K^{{\rm
soft}+{\rm sl}}_{\rm NLO}$ and $K^{{\rm soft}+{\rm sl}}_{\rm NNLO}$
for a proton-proton {\abbrev CMS} energy of (a)~14~TeV and (b)~2~TeV.
By comparing to 
\fig{fig::khksoft} one can see that the formally sub-leading terms clearly
dominate the cross section.
\Fig{fig::sigmasoftsl} shows the effect of the approximate sub-leading
terms on the total cross section. One should bear in mind that the agreement
of the solid and dotted lines in these two figures is due to the
dominance of $\bar\sigma_{gg}^{(2),\rm sl}$ which has
been added both to our $\hat\sigma_{gg}^{(2),\rm soft}$ and the
approximate  $\bar\sigma_{gg}^{(2),\rm soft}$.
The \nnlo\ corrections shown in
Figs.~\ref{fig::khksoftsl} and \ref{fig::sigmasoftsl} are quite large
and would indicate a very slow perturbative convergence
of the total cross section.  However, it is expected that the use of
{\abbrev NNLO} parton distributions, when they become available, will
moderate the size of the \nnlo\ terms.

\section{Conclusions}
The partonic cross section $gg\to H + X$ has been calculated to \nnlo\
in the limit of soft emission. The resulting hadronic Higgs production
rate exhibits well-behaved perturbative convergence properties.
However, the effects of soft emission are expected to be
significantly smaller than the formally non-leading contributions
$\propto \ln^i(1-x)$, for which an approximate result has been
obtained in ref.~\cite{gghresum}. This makes the complete evaluation
of the \nnlo\ corrections even more imperative.

\subsection*{Acknowledgments}
We would like to thank S.~Dawson and W.~Vogelsang for many helpful
discussions. The work of R.V.H. is supported by {\it Deutsche
Forschungsgemeinschaft} and that of W.B.K. by the United States
Department of Energy under grant DE-AC02-98CH10886.

\subsection*{Note added}
The subject of this paper has also been addressed by Catani, de
Florian and Grazzini~\cite{CFG}.  Their analytical results are in full
agreement with ours. Note, however, that their definition of the soft
limit differs from ours by a global factor of $x$ (cf. Eq.(6) of our
paper to Eq.(2) of \cite{CFG}). This means that in \cite{CFG} some of
the sub-leading terms (cf. Eqs.(32),(33)) are attributed to the soft
contribution, so that the numerical result for our soft limit is not
directly comparable to their ``SV'' approximation.  Other numerical
differences for the final results (our $\check\sigma_{gg}^{(i),{\rm
soft}+{\rm sl}}$, their $\sigma^{SVC}$) are due to the different sets
of parton distributions and the different treatment of unknown
sub-leading terms proportional to $\ln^2(1-x)$ and $\ln(1-x)$.


\def\app#1#2#3{{\it Act.~Phys.~Pol.~}{\bf B #1} (#2) #3}
\def\apa#1#2#3{{\it Act.~Phys.~Austr.~}{\bf#1} (#2) #3}
\def\annphys#1#2#3{{\it Ann.~Phys.~}{\bf #1} (#2) #3}
\def\cmp#1#2#3{{\it Comm.~Math.~Phys.~}{\bf #1} (#2) #3}
\def\cpc#1#2#3{{\it Comp.~Phys.~Commun.~}{\bf #1} (#2) #3}
\def\epjc#1#2#3{{\it Eur.\ Phys.\ J.\ }{\bf C #1} (#2) #3}
\def\fortp#1#2#3{{\it Fortschr.~Phys.~}{\bf#1} (#2) #3}
\def\ijmpc#1#2#3{{\it Int.~J.~Mod.~Phys.~}{\bf C #1} (#2) #3}
\def\ijmpa#1#2#3{{\it Int.~J.~Mod.~Phys.~}{\bf A #1} (#2) #3}
\def\jcp#1#2#3{{\it J.~Comp.~Phys.~}{\bf #1} (#2) #3}
\def\jetp#1#2#3{{\it JETP~Lett.~}{\bf #1} (#2) #3}
\def\jhep#1#2#3{{\it J.~High~Energy~Phys.~}{\bf #1} (#2) #3}
\def\mpl#1#2#3{{\it Mod.~Phys.~Lett.~}{\bf A #1} (#2) #3}
\def\nima#1#2#3{{\it Nucl.~Inst.~Meth.~}{\bf A #1} (#2) #3}
\def\npb#1#2#3{{\it Nucl.~Phys.~}{\bf B #1} (#2) #3}
\def\nca#1#2#3{{\it Nuovo~Cim.~}{\bf #1A} (#2) #3}
\def\plb#1#2#3{{\it Phys.~Lett.~}{\bf B #1} (#2) #3}
\def\prc#1#2#3{{\it Phys.~Reports }{\bf #1} (#2) #3}
\def\prd#1#2#3{{\it Phys.~Rev.~}{\bf D #1} (#2) #3}
\def\pR#1#2#3{{\it Phys.~Rev.~}{\bf #1} (#2) #3}
\def\prl#1#2#3{{\it Phys.~Rev.~Lett.~}{\bf #1} (#2) #3}
\def\pr#1#2#3{{\it Phys.~Reports }{\bf #1} (#2) #3}
\def\ptp#1#2#3{{\it Prog.~Theor.~Phys.~}{\bf #1} (#2) #3}
\def\ppnp#1#2#3{{\it Prog.~Part.~Nucl.~Phys.~}{\bf #1} (#2) #3}
\def\sovnp#1#2#3{{\it Sov.~J.~Nucl.~Phys.~}{\bf #1} (#2) #3}
\def\sovus#1#2#3{{\it Sov.~Phys.~Usp.~}{\bf #1} (#2) #3}
\def\tmf#1#2#3{{\it Teor.~Mat.~Fiz.~}{\bf #1} (#2) #3}
\def\tmp#1#2#3{{\it Theor.~Math.~Phys.~}{\bf #1} (#2) #3}
\def\yadfiz#1#2#3{{\it Yad.~Fiz.~}{\bf #1} (#2) #3}
\def\uspfiz#1#2#3{{\it Usp.~Fiz.~Nauk~}{\bf #1} (#2) #3}
\def\zpc#1#2#3{{\it Z.~Phys.~}{\bf C #1} (#2) #3}
\def\ibid#1#2#3{{ibid.~}{\bf #1} (#2) #3}


\end{document}